\documentclass[12pt]{article}  
\usepackage{a4,latexsym,amsmath}
\usepackage[latin1]{inputenc}
\usepackage[T1]{fontenc}
\usepackage{amsfonts}
\usepackage{dsfont}
\usepackage[doublespacing]{setspace}
\usepackage[footnotesize]{caption}
\usepackage{graphicx}
\usepackage{natbib}
\usepackage{booktabs}
\usepackage{float}
\usepackage{wasysym}
\usepackage{ulem}
\usepackage[
  bookmarks=false,
  pdfpagelabels=false,
  hyperfootnotes=false,
  hyperindex=false,
  pageanchor=false,
  colorlinks,
]{hyperref}

\usepackage{geometry}
\usepackage{siunitx}
\usepackage{fancyvrb}
\usepackage{verbatim}
\usepackage{xcolor}

 \usepackage{textcomp}
\usepackage[T1]{fontenc}

\makeatletter 
\let\saved@hyper@linkurl\hyper@linkurl 
\let\saved@hyper@link@\hyper@link@
\AtBeginDocument{%
  \NoHyper
  \let\hyper@linkurl\saved@hyper@linkurl  
  \let\hyper@link@\saved@hyper@link@  
}
\makeatother

\usepackage{abstract}

\usepackage{multirow}

\newcommand{\beginsupplement}{%
        \setcounter{table}{0}
        \renewcommand{\thetable}{C\arabic{table}}%
        \setcounter{figure}{0}
        \renewcommand{\thefigure}{C\arabic{figure}}%
         \setcounter{equation}{0}
        \renewcommand{\theequation}{B.\arabic{equation}}
     }

\newcommand{\mat}[1]{\boldsymbol{#1}}

\begin{document}
\normalem

\title{
  \textbf{Regularization and Model Selection for Ordinal-on-Ordinal Regression with Applications to Food Products' Testing and Survey Data
}
} 

\author{
  Aisouda Hoshiyar\thanks{To whom correspondence should be addressed:
 \texttt{aisouda.hoshiyar@hsu-hh.de}} \thanks{School of Economics and Social Sciences, Helmut Schmidt University, Hamburg, Germany},\ Laura H. Gertheiss\thanks{Department of Agricultural Economics and Rural Development, Georg August University, G\"ottingen, Germany
},\ \& Jan Gertheiss\footnotemark[2]}

\maketitle

\begin{center}
\textbf{Abstract}
\end{center}
Ordinal data are quite common in applied statistics. Although some model selection and regularization techniques for categorical predictors and ordinal response models have been developed over the past few years, less work has been done concerning ordinal-on-ordinal regression. Motivated by a consumer test and a survey on the willingness to pay for luxury food products consisting of Likert-type items, we propose a strategy for smoothing and selecting ordinally scaled predictors in the cumulative logit model. First, the group lasso is modified by the use of difference penalties on neighboring dummy coefficients, thus taking into account the predictors' ordinal structure. Second, a fused lasso-type penalty is presented for the fusion of predictor categories and factor selection. The performance of both approaches is evaluated in simulation studies and on real-world data.

\vspace{1cm}


\section{Introduction}
  
Ordered categorical variables arise commonly in regression modeling. Still, proper treatment of the underlying scale level and appropriate statistical modeling is an issue of discussion.
In many applications, particularly the nature of ordinal predictors such as rating scales is neglected, and a classical linear model is fit instead, thus assuming a metric scale level~\citep{Tutz:2014}. On the other hand, statistics textbooks often transform ordinal covariates into dummy variables, thus using the variables' nominal scale level only. Ignoring that categories are ordered implies, however, that we do not make full use of all the information contained in the data. Sometimes, when using dummy-coded ordinal predictors, monotonicity in the corresponding regression parameters is assumed to exploit the categories' ordering~\citep[compare, e.g.,][]{Young:1976, LinMeuKooGro:2007, Rufibach:2010}, which is also known under the name `isotonic regression'~\citep{BarlowEtal:1972}, but the assumption of monotonicity is not always plausible \citep{LinMeuKooGro:2007}. If the dependent variable is ordinal, ordered response models as treated in most advanced textbooks on statistical modeling~\citep[e.g.,][]{Fahrmeiret13} can be used. These models are typically recommended in the statistics literature but far less used in applications; compare, e.g., \citet{ChrBro:2013}. \citet{Kruschke:2018}, for instance, demonstrated that applying metric models to ordinal data can lead to significant errors, such as Type I and Type II errors, and argued for the use of ordered-probit models to ensure more accurate and reliable analysis. 
Although modeling of ordinal response variables has been well investigated, ordinal-on-ordinal regression is hardly mentioned. To investigate the relationship between two ordinal variables, basically, any statistical tool designed for analyzing the relationship between two categorical variables arranged in a contingency table could be used. Those tools include so-called association models. For instance, the RC model~\citep{Goodman:1979} is a type of association model that allows us to study the interaction between the row and column categories. Such models are particularly useful in understanding how different categories of one variable are associated with the categories of another variable. However, when the number of covariates and/or the number of levels per variable becomes large, RC models are no longer practically manageable. In this work, we hence focus on ordinal regression models to analyze the effects of a set of ordinal explanatory variables on an ordinal response variable. 
To give an example of such a setting, let us consider a survey on luxury food~\citep{Hartmann:Significance, Hartmann:Segmentation}, which will be the primary case study here. This survey collected data on various items measured on a 5-point Likert-type scale. The primary goal of the original study was to segment German consumers based on their perceived dimensions of luxury food and the shift of consumer consumption motives toward indulgence, quality, and sustainability. In what follows, we will focus on a regression problem where the response variable is the willingness to pay for luxury food products, which is also evaluated on an ordinal 5-point scale. The set of predictors considered contains 43 ordinally scaled statements (items) on eating habits, shopping places/habits, the importance of price, eating style, and the subjective definition of luxury food. In high-dimensional settings like this, however, with ordinal data both on the left and right-hand side of the regression, the full regression model, such as a cumulative logit model, is often hard to fit by common maximum likelihood. Penalty-based regularization has been shown to be a viable means to handle such situations, and various regularization techniques have been proposed for regression models with categorical variables over the last two decades. \citet{Meier:2008}, for instance, used the group lasso penalty \citep{Yuan:2006} for factor selection in logistic regression. \citet{Tutz:2014, Tutz:2016} proposed methods for the selection of ordinally scaled independent variables in the classical linear, the logistic, and the log-linear (Poisson) model. Regularized ordinal regression has been considered by \citet{ArcWil:2012} and \citet{Wurm:2021} using a lasso-type \citep{Tibshirani:1996} and elastic net~\citep{ZouHas:2005} penalty, respectively.  

The limitation of the penalty approaches above, however, is that they assume either the response or the predictors to be of metric scale or binary. In this article, we propose a penalty-based strategy for model selection with ordinally scaled  covariates and smoothing across covariate levels in ordinal regression, specifically, the cumulative logit model. For doing so, the original group lasso is modified by the use of a difference penalty on neighboring dummy coefficients, thus taking into account their ordinal structure. As an alternative, the use of a fused lasso-type penalty~\citep{Tibshirani:2005} will be discussed.  

The remainder of the paper is organized as follows. Section~\ref{sec2} introduces two motivating data examples on food products and the corresponding modeling framework. In Section~\ref{sec3}, we will describe in detail the two penalty methods for ordinal predictors mentioned above. Some illustrative simulation studies in Section~\ref{sec4} examine the quality of the proposed methodology. In Section~\ref{sec5}, the proposed methods are applied to data from a consumer test and the study on luxury food expenditure. In Section~\ref{sec6}, we conclude with a summary and discussion. All computations were done using the statistical program R~\citep{R}. The proposed methods are implemented in the open-source R add-on package \texttt{ordPens} \citep{ordPens:2021} available on CRAN and Github (\url{https://github.com/ahoshiyar/ordPens}). 

\section{Data Examples and Modeling Framework}\label{sec2}

\subsection{Two Case Studies}\label{sec2:cs}
\subsubsection{Perception and Acceptance of Boar-tainted Meat}\label{boart}

\begin{figure}[!ht]\centering
 \includegraphics[width=\linewidth]{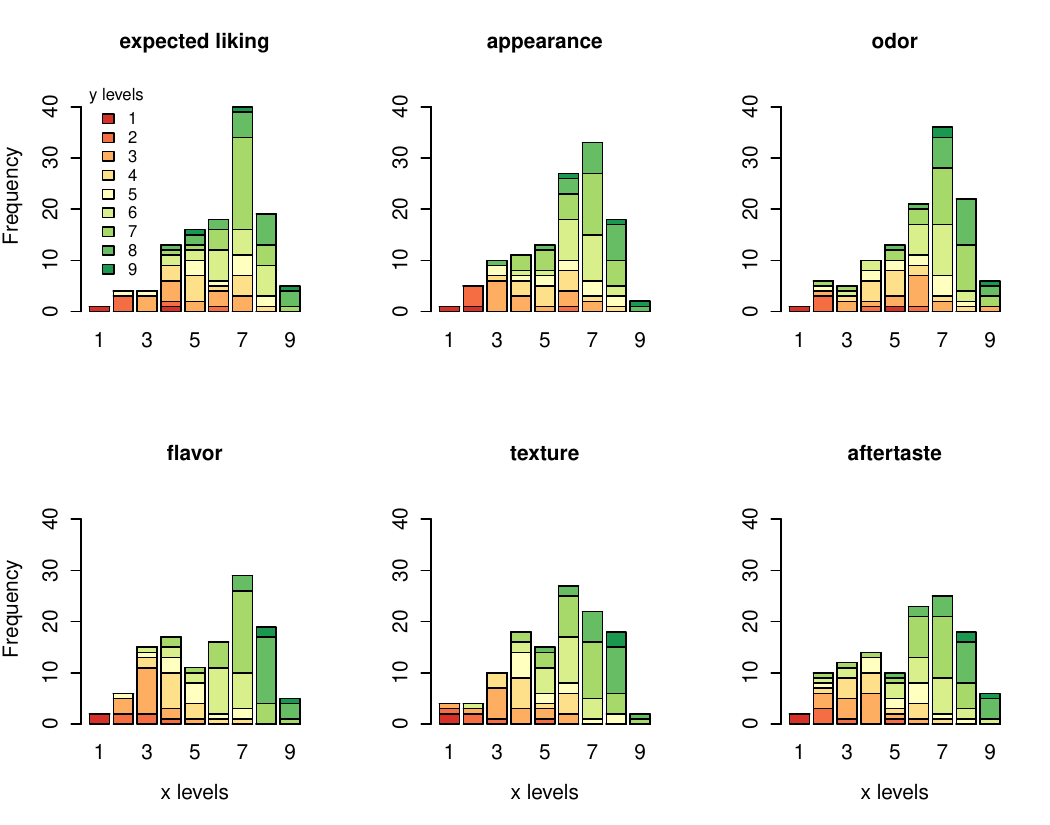} 
\caption{\label{hoshiyar:fig2bars} 
Illustration of the sensory data. Frequency distribution for the covariates with levels $1 = \text{`dislike extremely'}$ to $9 = \text{`like extremely'}$, segmented by frequencies of response $y$ (`overall liking'), with colors corresponding to response categories $1 = \text{`dislike extremely'}$ to $9 = \text{`like extremely'}$ (see top left).}
\end{figure}

For illustrative purposes, we first consider a consumer test about the perception and acceptance of boiled sausages from strongly boar-tainted meat, conducted by the Department of Animal Sciences, University of G\"ottingen \citep{Meier:2016}. 
Due to the European call for alternatives to surgical castration of male piglets, as this practice was proven to be cruel \citep{EU, EU2}, there is rising interest in the production and processing of boar meat. However, there is a disadvantageous risk that off-flavors occur, the so-called boar taint, which is to be dealt with. An effort is being put into improving the processing of tainted raw material in a sustainable way with the primary aim of making it `fit for human consumption'. Specifically, one challenge lies in masking the undesirable odor --and the associated flavor-- by combining spices/herbs/aromas and developing meat products that are palatable and accepted by consumers.                 
The study at hand examined six sausages of different types (raw smoked-fermented or boiled emulsion-type sausages) with varying proportions of tainted raw material ($50\%$, $100\%$, plus a control product without tainted material). 
120 participants were recruited to consume and evaluate the products with regard to different features; see \citet{Meier:2016} for details on the recruited consumer sample.
For illustrational purposes, we focus on boiled sausages here with a proportion of $50\%$ tainted boar meat (product $6$).
In order to process highly tainted meat sustainably, we aim to detect the relevant factors for the consumers' overall liking. The six ordinal predictors considered {\it `expected liking', `appearance', `odor', `flavor', `texture', `aftertaste'}, and the ordinal response {\it `overall liking'} are measured on a $9$-point scale ($1 = \text{`dislike extremely'}$ to $9 = \text{`like extremely'}$). The data is summarized through barplots in Figure~\ref{hoshiyar:fig2bars}. Using a regression model, the effect of the ordinal covariates on the ordinal factor {\it overall liking} is to be investigated.  

\subsubsection{Consumers' Willingness to Pay for Luxury Food}\label{luxury}

As a second and main case study, we consider the before-mentioned survey concerning, among other things, the consumers' willingness to pay for `luxury food'~\citep{Hartmann:Significance, Hartmann:Segmentation}. The data was collected by a commercial panel provider and is representative of the German population with respect to age, gender, and education \citep[see][for details]{Hartmann:Significance, Hartmann:Segmentation}. The part of the dataset we examine consists of 821 observations of 44 Likert-type items on personal luxury food definitions, eating and shopping habits, diet styles, and food price sensitivities. One such item is, for example, ``I associate a high price in food with particularly good quality'', with coding scheme $-2 = $ `strongly disagree', $\ldots \,$, $+2 = $ `fully agree'. The coding scheme of the remaining items is similar and given in Table~C3 in the online supplements. Our response of interest is whether participants would be willing to pay a higher price for a food product that they associate with luxury, measured again on the ordinal $-2$ to $+2$ scale. The subset of the data that is analyzed here has been made publicly available at \url{https://zenodo.org/record/8383248}~\citep{LuxDat}.  
A different subset of the data has previously been studied by \citet{Tutz:2016}, but their focus was on aspects regarding more general behavior when buying and consuming food products. Also, their response of interest was different, namely the (approximate) household's weekly expenditure on food, assumed to be measured on a metric scale.

\subsection{Ordinal-on-Ordinal Regression}\label{ioi}
 
The cumulative model~\citep{Mccullagh:1980} is probably the most popular regression model for ordinal response variables; compare, e.g., \citet{Agresti:2010, Agresti:2013}, \citet{Fahrmeir:2001}, \citet{Fahrmeiret13}, \citet{Tutz:2022}. It can be motivated such that the observable response variable $y$ $\in \{1,\ldots,c\}$ is a categorized version of a latent continuous variable $u$.  
The link between the response variable $y_i$ (i.e., $y$ for subject $i= 1,\ldots, n$) and the corresponding latent variable $u_i$ is then defined by the threshold mechanism
$$
y_i = r \Longleftrightarrow \theta_{r-1} < u_i \leq \theta_{r}, \quad r=1,\ldots,c,$$
where $-\infty = \theta_0 < \theta_1 < \ldots < \theta_{c} = \infty$ are the ordered thresholds.
Given $p$ numeric covariates, the latent variable is typically modeled as 
$
u_i = \mat{x}^\top_i \mat{\beta} + \epsilon_i, 
$
where $\mat{x}_i = (x_{i1},\ldots,x_{ip})^\top$ denotes the covariate vector for subject $i$, $\mat{\beta} = (\beta_1,\ldots,\beta_p)^\top$ is the parameter vector and $\epsilon_i$ is an error variable with distribution function $F$.  
From these assumptions, it follows that 
$$
P(y_i \leq r) = P(u_i \leq \theta_r) = F(\theta_r - \mat{x}^\top_i \mat{\beta}) = F(\eta_{ir}),
$$  
with $\eta_{ir}= \theta_r - \mat{x}^\top_i \mat{\beta}$ being the linear predictor.
Common choices for $F(\cdot)$ are the logistic and the standard normal cumulative distribution function, resulting in the cumulative logit and the cumulative probit model, respectively.  
The most widely used cumulative model is the cumulative logit model \citep[cf.][]{Tutz:2022}, which we will concentrate on here. The cumulative logit model is also known as the proportional odds model (POM) since the ratio of the cumulative odds for two populations is postulated to be the same across all (response) categories.
 
With ordinal covariates, without loss of generality, each $\mat{x}_i = (x_{i1},\ldots,x_{ip})^\top$ contains integers, with entry $x_{ij}$ indicating the level of the $j$th variable that is observed at the $i$th subject. So let the $j$th variable have (potential) values $1,\ldots,k_j$ and define dummy variables such that $\mat{z}_{ijl}=1$ if $x_{ij}=l$, $l=1,\ldots,k_j$, and $\mat{z}_{ijl}=0$ otherwise. 
The model to be fitted then has the linear predictor 
\begin{equation}\label{eta}
\eta_{ir} = \theta_r - \sum_{j=1}^p \sum_{l=1}^{k_j} z_{ijl} \beta_{jl} = \theta_r - \sum_{j=1}^p \mat{z}_{ij}^\top \mat{\beta}_j,
\end{equation} 
where vector $\mat{z}_{ij} = (z_{ij1},\ldots,z_{ijk_j})^\top$ collects the dummy variables belonging to the $j$th predictor variable, and $\mat{\beta}_j = (\beta_{j1},\ldots,\beta_{jk_j})^\top$ contains the corresponding regression coefficients. Estimation of unknown coefficients and thresholds $\theta_r$ relies on well-known maximum likelihood principles. Further details can be found in the online appendix (Section A) and in \citet{Mccullagh:1980}.
For reasons of identifiability, however, some restrictions are needed. For instance, by choosing a so-called reference category for each covariate and setting the corresponding regression coefficient to zero. Alternatively, one may set $\sum_l \beta_{jl} = 0 \ \forall j$, which is also known as `effect coding', and will be used throughout this paper. With respect to parameter estimates, both restrictions are equivalent.

\begin{figure}[!ht]\centering
 \includegraphics[width=\linewidth]{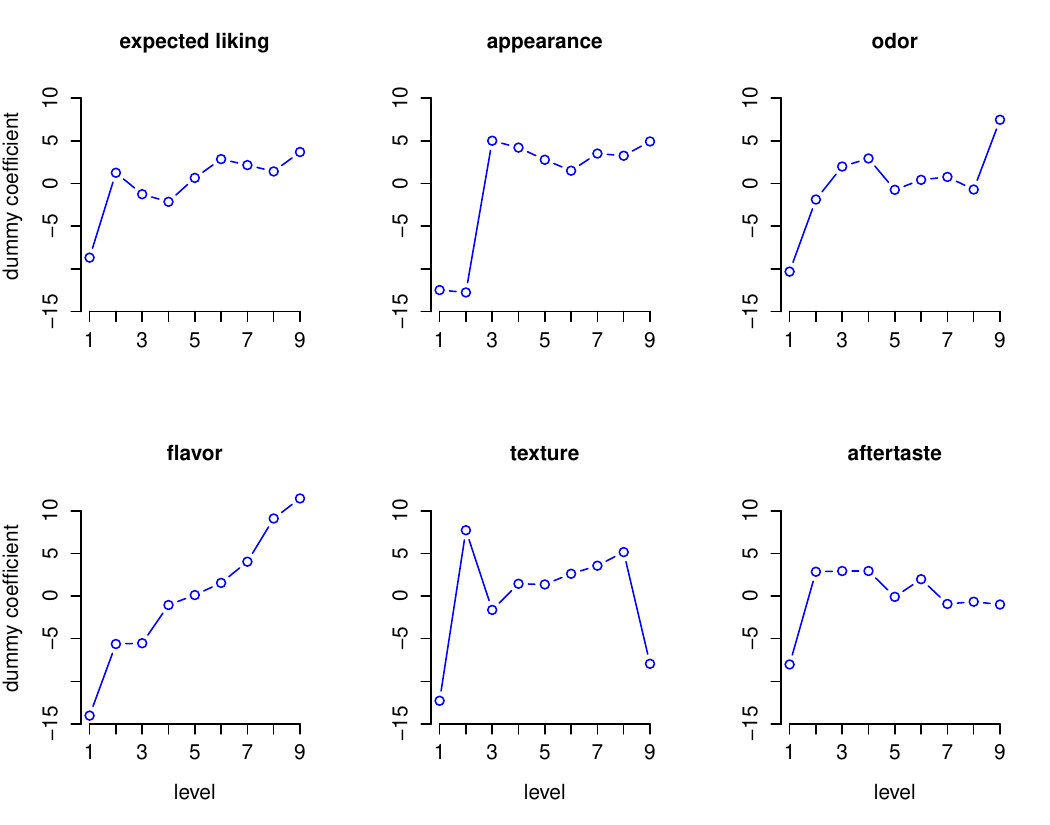} 
\caption{\label{hoshiyar:fig2} 
\texttt{polr} estimates of the regression coefficients as functions of class labels for the sensory data from Section~\ref{boart}. The estimated thresholds are: $-34.22$, $-1.20$,  $4.84$, $7.32$, $8.86$, $11.29$, $15.75$, $21.98$. }
\end{figure}

For illustration, Figure~\ref{hoshiyar:fig2} gives the estimated dummy coefficients in the proportional odds model when fit to the sensory data from Section~\ref{boart} using the \texttt{MASS::polr} function in R \citep{Venables:2002}. In the first attempt to estimate the model, however, the algorithm did not converge. The function failed to find suitable starting values. 
This is a known problem with the function \texttt{polr} since it estimates a (binary) logit model to initiate, and the function terminates when the GLM does not converge. Convergence problems occur when the fitted probabilities are extremely close to zero or one, as also addressed by \citet[p. 197-198]{Venables:2002}.
However, fitting was possible when inserting some group lasso penalized estimates as starting values (compare Section~\ref{sec3}). It is seen that \texttt{polr} estimates are very wiggly and thus hard to interpret. Moreover, some effects appear to be extremely large (note that the effects refer to the logit scale here), causing some fitted response probabilities to be numerically zero or one. The reason for that becomes clear when looking at the data in Figure~\ref{hoshiyar:fig2bars}. Typically, there is only a small number of samples found in the lowest x-category (`1'), with most (or even all) observations falling in the same and most extreme y-category `1', whereas y-level `1' is hardly found for x-levels greater than 1. As a consequence, fitting by pure maximum likelihood leads to (seemingly) extreme effects. At this point, it should be noted that \texttt{polr} is not the only routine available to fit cumulative logit/proportional odds models. For instance, the R package \texttt{ordinal}~\citep{Christensen:2023} could be used as well. However, this package also uses common maximum likelihood, and results are virtually identical to those obtained with \texttt{polr}. The main advantage of the \texttt{ordinal} package over \texttt{polr} is that it also allows to include random effects, which is often needed in sensometrics but not the focus of this paper.

\section{Regularization and Model Selection}\label{sec3} 
 
\subsection{Groupwise Selection and Smoothing}\label{ORS}

With categorical predictors included in the model, resulting in a large (potentially enormous) number of parameters to be estimated, typically, two challenges arise. The first step is to decide on the variables to include in the regression model. Secondly, we have to decide which categories within one categorical predictor should be distinguished.   
Furthermore, if maximizing the likelihood using (\ref{eta})is maximized, only the nominal scale level of the predictor variables is used. As already discussed by \cite{Tutz:2014, Tutz:2016} for other types of regression models (namely, models with numerical, one-dimensional response), it can be beneficial to use special penalties to incorporate the predictors' ordinal scale level. In particular, penalizing differences between coefficients that belong to adjacent categories in combination with a group lasso as done by \citet{GerHogObeTut:2011} seems promising. The rationale behind this approach is as follows. If the levels of a categorical predictor are ordered, it may be assumed that the (expected) response does not change drastically but rather smoothly between neighboring predictor categories. For simulation studies showing that this method is highly competitive in the classical linear model both in terms of variable selection and estimation accuracy, see \citet{Tutz:2014} and \citet{GerTut:2023}. To implement the approach in the cumulative logit model, we will extend the logistic group lasso estimator for binary responses \citep{Meier:2008} to the class of cumulative logit models. This type of estimator is given by the minimizer of the function
\begin{equation}\label{penlogl} 
l_{\lambda}(\mat{\theta},\mat{\beta}) = - \frac{1}{n} l(\mat{\theta},\mat{\beta}) + \lambda \sum_{j=1}^p J_j(\mat{\beta}_j),
\end{equation}
where $l(\mat{\theta},\mat{\beta})$ is the log-likelihood of the cumulative logistic distributionin (2.2); vectors $\mat{\theta}$ and $\mat{\beta}$ collect all thresholds $\theta_1,\ldots,\theta_{c-1}$ and subvectors $\mat{\beta}_1,\ldots,\mat{\beta}_p$, respectively.  
In order to take into account the ordinal structure of the predictors, we modify the usual $L_2$-norm of the group lasso by the first-order difference penalty functions
\begin{equation}\label{defJ}
J_j(\mat{\beta}_j) = \sqrt{\Bigg\{ \sum_{l=2}^{k_j} \text{df}_j (\beta_{jl} - \beta_{j,l-1})^2\Bigg\}} =   \sqrt{\text{df}_j} ||\mat{D}_{1,j} \mat{\beta}_j ||_2,
\end{equation} 
with $k_j$ being the number of levels of the $j$th covariate and $\text{df}_j=k_j-1$. 
$||\cdot||_2$ denotes the $L_2$-norm and $\mat{D}_{1,j}$ is the matrix that produces differences of first order for the $j$th predictor and is given in Appendix~A.  
This first-order penalty both enforces the selection of the whole group of parameters that belong to the same categorical predictor and simultaneously smoothes over the ordered categories.

To find the final solution, we can utilize the block coordinate gradient descent method of \citet{Tseng:2009}, as described in \citet{Meier:2008}. Namely, 
combining a quadratic approximation of the log-likelihood with an additional line search.
Based on the coordinate descent algorithm, the R package \texttt{grplasso}~\citep{Meier:2020} provides (standard) group lasso estimation for the Poisson, the logistic, and the classical linear model. 
For penalized ordinal-on-ordinal regression in terms of a cumulative logit model with ordinal predictors and corresponding ordinal smoothing penalty~\eqref{defJ} some modifications have to be made. 

Using a second-order Taylor series expansion at $\hat{\mat{\gamma}}^{[t]} = (\hat{\mat{\theta}}^{[t]\top}, \hat{\mat{\beta}}^{[t]\top})^\top$, i.e., at the current estimate in iteration $t$ of the algorithm, and replacing the Hessian of $l(\cdot)$ by some suitable block diagonal matrix $\mat{H}^{[t]} = \text{diag}(\mat{H}_0^{[t]},\mat{H}_1^{[t]},\ldots,\mat{H}_p^{[t]})$, we specify 
\begin{equation} \label{M}
\begin{split}
 M_\lambda^{[t]}(\textbf{\text{d}}) &= - \Bigl\{ l(\hat{\mat{\gamma}}^{[t]}) + \textbf{\text{d}}^\top \nabla l(\hat{\mat{\gamma}}^{[t]}) + \frac{1}{2} \textbf{\text{d}}^\top \mat{H}^{[t]} \textbf{\text{d}} \Bigr\} + \lambda \sum_{j=1}^p \sqrt{\text{df}_j}|| \mat{D}_{1,j} \hat{\mat{\beta}}_{j}^{[t]} + \textbf{\text{d}}_{j}  ||_2  \\
&\approx l_\lambda(\hat{\mat{\gamma}}^{[t]}+\textbf{\text{d}}),
\end{split}
\end{equation}
with $\textbf{\text{d}} = (\textbf{\text{d}}_0^\top,\textbf{\text{d}}_1^\top,\ldots,\textbf{\text{d}}_p^\top)^\top \in \mathbb{R}^{c-1 + \sum_j(k_j - 1)}$, where the index 0 denotes the elements belonging to $\theta_1,\ldots,\theta_{c-1}$, and $\mat{D}_{1,j}$ from above (also see Appendix A.3).  
Minimization of $M_\lambda^{[t]}(\textbf{\text{d}})$ is carried out with respect to the $j$th parameter group, $j=0,1,\ldots,p$, as follows: We use the ($k_j \times k_j$) submatrix $\mat{H}_j^{[t]}=h_j^{[t]}\mat{I}_{k_j}$ for some scalar $h_j^{[t]} \in \mathbb{R}$, where a possible choice is 
$$
h_j^{[t]} = \min\{\max\{\text{diag} ( \nabla^2 l(\hat{\mat{\gamma}}^{[t]})_{jj} ), 10^{-6} \}, 10^8 \},
$$ 
and $\mat{I}_{k_j}$ is the identity matrix of corresponding dimensions. 

If 
\begin{equation}\label{cond}
|| \nabla  l(\hat{\mat{\gamma}}^{[t]})_{j} - h_j^{[t]} \hat{\mat{\beta}}^{[t]}_j  ||_2 \leq \lambda \sqrt{\text{df}_j},
\end{equation}
the minimizer of $M_\lambda^{[t]}(\textbf{\text{d}})$ in (\ref{M}) w.r.t. to group $j$ is
\begin{equation*} 
\textbf{\text{d}}^{[t]}_j = -\hat{\mat{\beta}}_j^{[t]}.
\end{equation*} 
Otherwise
\begin{equation*}
\textbf{\text{d}}^{[t]}_j = - \frac{1}{h_j^{[t]}} \Bigg\{ \nabla  l(\hat{\mat{\gamma}}^{[t]})_{j}  - \lambda \sqrt{\text{df}_j} \frac{  \nabla  l(\hat{\mat{\gamma}}^{[t]})_{j} - h_j^{[t]} \hat{\mat{\beta}}_j^{[t]} }{ || \nabla  l(\hat{\mat{\gamma}}^{[t]})_{j} - h_j^{[t]} \hat{\mat{\beta}}_j^{[t]} ||_2  }   \Bigg\}.
\end{equation*}
If $\textbf{\text{d}}^{[t]} \neq \mat{0}$, an (inexact) line search using the Armijo rule must be performed. Let $\alpha^{[t]}$ be the largest value among the grid $ \{\alpha_o \delta^l\}_{l\geq0} $ s.t. 
\begin{equation*}
l_\lambda(\hat{\mat{\gamma}}^{[t]}+\alpha^{[t]}\textbf{\text{d}}^{[t]}) - l_\lambda(\hat{\mat{\gamma}}^{[t]}) \leq \alpha^{[t]} \sigma \Delta^{[t]}, 
\end{equation*}
where $0<\sigma<1$, $\alpha_0>0$ and $\Delta^{[t]}$ is the improvement in $l_\lambda(\cdot)$
when using a linear approximation for the objective function
\begin{equation*}
\Delta^{[t]} = -\textbf{\text{d}}^{[t]\top} \nabla  l(\hat{\mat{\gamma}}^{[t]})  + \lambda \sqrt{\text{df}_j} || \mat{D}_{1,j} \hat{\mat{\beta}}_j^{[t]} + \textbf{\text{d}}_j^{[t]} ||_2 - \lambda \sqrt{\text{df}_j} || \mat{D}_{1,j}  \hat{\mat{\beta}}_j^{[t]} ||_2.
\end{equation*}
Standard choices are  $\alpha_0 = 1$, $\delta = 0.5$ and $\sigma = 0.1$.
Finally, we specify
\begin{equation*}
\hat{\mat{\beta}}^{[t+1]} = \hat{\mat{\beta}}^{[t]} + \alpha^{[t]} \textbf{\text{d}}^{[t]}.
\end{equation*}
	\begin{singlespace} 
The described procedure, namely 
\begin{itemize}
\item looping through the block coordinates $j = 0,1,\ldots,p$:  
\item update $\mat{H}_{j}^{[t]} \leftarrow h_j^{[t]}\mat{I}_{k_j}$  
\item $\textbf{\text{d}}^{[t]} \leftarrow \text{arg min} \{M_\lambda^{[t]}(\textbf{\text{d}})\}$  
\item if $\textbf{\text{d}}^{[t]}\neq \mat{0}$ \\
 $\alpha^{[t]} \leftarrow$ line search \\
 $\mat{\beta}^{[t+1]} \leftarrow \mat{\beta}^{[t]} + \alpha^{[t]} \textbf{\text{d}}^{[t]}$
\end{itemize} 
is repeated until some convergence criterion is met.
	\end{singlespace} 

Note that the check for differentiability in condition~(\ref{cond}) is not necessary for the intercept parameters (because those parameters are not penalized), i.e., $j=0$, and the solution can be directly computed as
$$
\textbf{\text{d}}_0^\top = - \frac{1}{h_0^{[t]}} \nabla l(\hat{\mat{\gamma}}^{[t]})_0.
$$ 

The strength of penalization is controlled by the parameter $\lambda$. With $\lambda = 0$, unpenalized maximum likelihood estimates for categorical variables as described in Section~\ref{ioi} are obtained.
If $\lambda \to \infty$ one obtains the extreme case that coefficients that
belong to the same predictor are estimated to be equal, which means that the corresponding predictor has no effect on the response. In fact, with any of the restrictions (reference category set to zero or effect coding), all parameters except for thresholds are estimated to be zero if $\lambda \to \infty$. Further, note that penalized estimates as proposed here are invariant against the concrete choice of the restrictions because the values of both the likelihood and penalty at~(\ref{defJ}) do not depend on the restriction chosen. 
Note that we do not penalize thresholds $\theta_r$.

\subsection{Ordinal-on-Ordinal Fusion}

If a variable is selected by the smoothed group lasso (\ref{defJ}), all coefficients belonging to that selected covariate are estimated as differing (at least slightly). However, it might be useful to fuse certain categories. Fusion of dummy coefficients of ordinal predictors has already been discussed by \citet{Tutz:2014, Tutz:2016} for regression models with numerical, one-dimensional response; namely, the classical linear model, and a generalized linear model with binary or count data (Poisson) response. Having data like those described in Section~\ref{sec2:cs} in mind, we extend this previous work to the multivariate case of multi-categorical, ordinal response models; specifically, the framework of the cumulative logit model. 
Fusion is done by a fused lasso-type penalty~\citep{Tibshirani:2005} using the $L_1$-norm on adjacent categories. That is, adjacent categories may end up with exactly the same coefficient values and one obtains clusters of categories.
The idea is to maximize $l_\lambda(\mat{\theta},\mat{\beta})$ from~(\ref{penlogl}) above
with
\begin{equation}\label{fus}
J_j(\mat{\beta}_j) =  \sum_{l=2}^{k_j} | \beta_{jl} - \beta_{j,l-1} |  .
\end{equation}

Using the $L_1$-norm for neighboring categories encourages that adjacent parameters are set exactly equal rather than just being close (as done by penalty (\ref{defJ})). Penalty (\ref{fus}) thus has the effect that neighboring categories may be fused (namely, if they have the same $\mat{\beta}$ values). 
The fused lasso also enables the selection of predictors: a predictor is excluded if all its categories are combined into one cluster. As already pointed out in Section~\ref{ORS} above, any of the restrictions considered here then means that all parameters belonging to the respective predictor are set to zero. As before, penalty~(\ref{fus}) is invariant against the concrete choice of restriction. After reparametrization, our fused lasso variant can be fit using existing software for the lasso. Technical details can be found in Appendix~A.3.

\subsection{Tuning parameter, model and feature selection} 
\subsubsection{Cross Validation}\label{boarcv}

In the definitions and algorithms above, the penalty parameter $\lambda$ was fixed at some specific value. However, the choice of an optimal value for $\lambda$ should be made using the data at hand. A common strategy in penalized regression is $K$-fold cross-validation as described in many textbooks, e.g.,~\citet{HasTibFri:2009}.
As a measure of performance, we calculate the \textit{Brier Score} \citep{Brier:1950}
$ \text{BS} = \sum_i \sum_r (v_{ir} - \pi_{ir})^2$, with the $c$-dimensional vector $\mat{v}_i = (v_{i1},\ldots, v_{ic})^\top$ of indicator variables, where $v_{ir} = 1$ if $y_i = r$, and zero otherwise.
Now, over a fine grid $G$ of sensible values $\lambda \in G$, the optimal $\lambda$ can be determined by minimizing the cross-validated Brier Score. An alternative measure of performance is the so-called ranked probability score~\citep[see, e.g.,][]{Epstein:1969}. On the data considered in this paper, however, the results concerning the chosen $\lambda$'s were virtually identical.

\subsubsection{Stability Selection}

The previous section described a potential process for selecting the regularization parameter, with the selection of this parameter also implying the particular set of features being selected. Particularly for high-dimensional data, cross-validation techniques can be quite challenging as cross-validated choices may include too many variables \citep{Meinshausen:2006, Leng:2006}. If we are mainly interested in feature selection, we can alternatively apply stability selection as suggested by \citet{Meinshausen:2010}, a promising subsampling strategy combined with high-dimensional variable selection. In general, instead of selecting/fitting one model, the data are perturbed or subsampled many times, and we choose those variables that occur in a large fraction of runs. Specifically, for the $j$th variable, the estimated probability $\hat{\pi}^\lambda_j$ of being in the stable selection set corresponds to the frequency of being chosen over all subsamples for a specific $\lambda$. Or, in other words, we keep variables with a high selection probability $\hat{\pi}^\lambda_j \ge \pi_\text{thr}$ and neglect those with low selection probability. The cutoff value can therefore be seen as a tuning parameter, and a typical choice is $\pi_\text{thr} \in (0.6, 0.9)$ \citep{Meinshausen:2010}.  
Motivated by the concept of regularization paths, we draw the stability path: the probability for each variable to be selected when randomly subsampling from the data as a function of $\lambda$.   
\citet{Meinshausen:2010} emphasize that choosing the proper regularization parameter is much less critical for the stability path, which increases the chance of selecting truly relevant variables.

\section{Numerical Experiments}\label{sec4}  

\subsection{Study Design}

\begin{figure}[!ht]\centering
\includegraphics[width=0.85\linewidth]{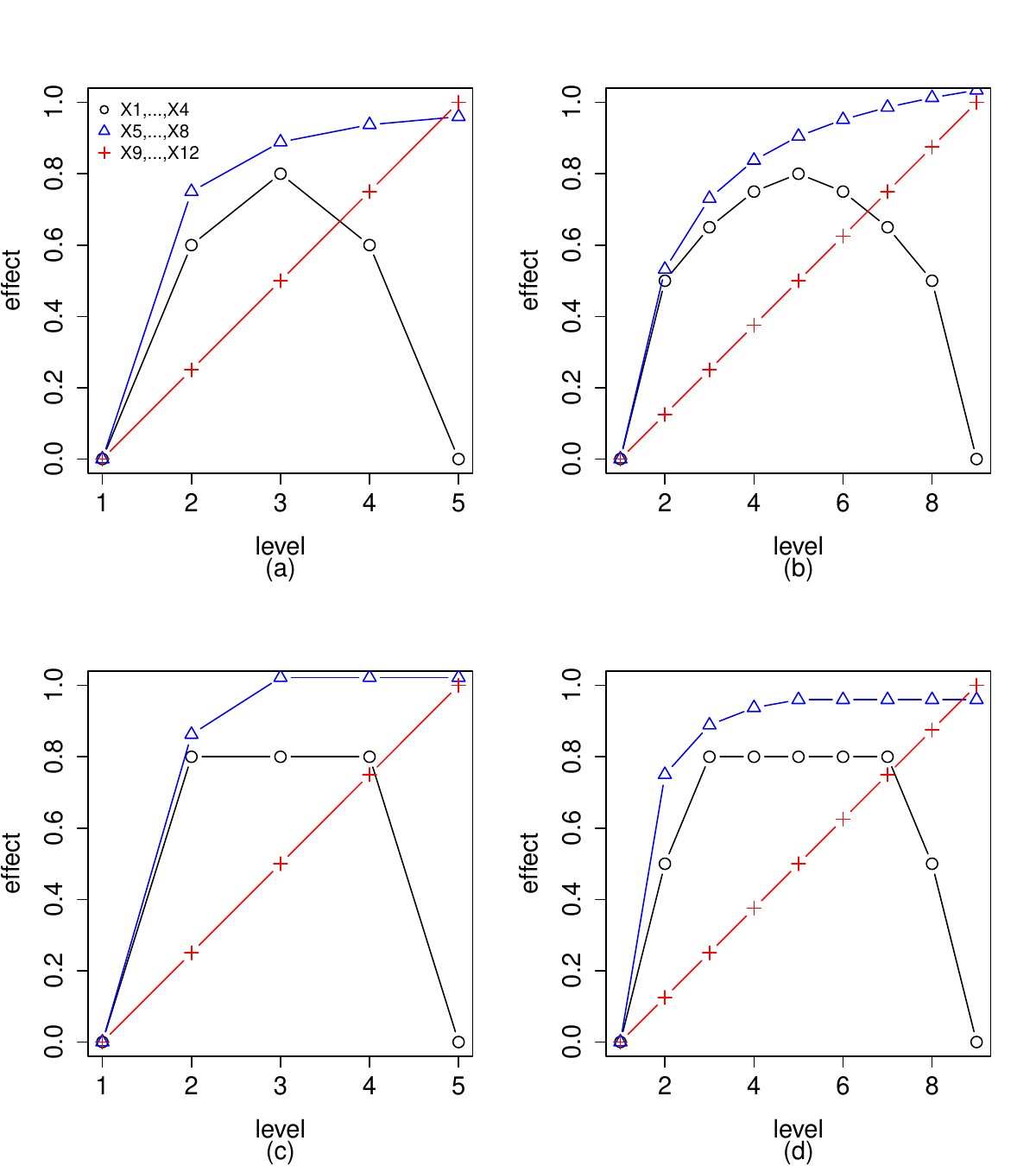} 
\caption{\label{hoshiyar:fig6a} True effects of influential predictors in the simulation study. (a) $5$ levels without fused effects; (b) $9$ levels without fused effects; (c) $5$ levels with fused effects; (d) $9$ levels with fused effects.}
\end{figure}

Before applying the presented methodology to the real-world data in Section~\ref{sec5}, we carry out a simulation study to investigate the properties of the proposed ordinal-on-ordinal selection approach using penalty (\ref{defJ}) and (\ref{fus}), respectively. 
We generate $p=50$ ordinally scaled covariates $X_1,\ldots,X_{50}$, including $38$ noise variables as follows: $X_1,\ldots,X_4$ have non-monotone effects, $X_5,\ldots,X_8$ are monotone but non-linear and the effect of $X_9,\ldots,X_{12}$ is linear across categories. The remaining $38$ predictors are irrelevant, i.e., with zero effects. Each predictor is assumed to have the same number of levels, either five or nine, depending on the concrete setting considered. Factor levels are then randomly drawn from $\{1,...,5\}$ or $\{1,...,9\}$, respectively. The (true) covariate effects are shown in Figure~\ref{hoshiyar:fig6a}. Using those predictors, we construct the ordinal response (with five levels) through the cumulative logit model with linear predictor~\eqref{eta} and thresholds ${\mat\theta} = (5.5,6.5,7.5,8.5)^\top$, and consider three different sample sizes $n=200,500,1000$. 

We assign ranks to the predictor items according to the order of non-zero ordinal group lasso coefficients in the coefficient path and call this ordinal rank selection (ORS). 
Similarly, we proceed with the non-zero ordinal fused lasso coefficients in the corresponding coefficient path and call this ordinal rank fusion (ORF). %
We compare these two approaches to two competitors: (1) the standard proportional odds model (POM) fit by \texttt{polr} and combined with AIC-based forward stepwise selection where the ordinal covariate is treated as nominal, i.e., it is dummy-coded without any penalization, and (2) treating the predictor as numeric and employing \texttt{ordinalNet} with a standard lasso penalty for selection. 
Note that the usual \texttt{ordinalNet} function offers no (nominal/ordinal) group lasso option. Thus, the selection of ordinal responses is only accessible if we ignore the categorical nature of the covariates.  
The entire data generation and estimation/selection process was repeated 100 times. The results are discussed below.

\subsection{Results} 

\subsubsection{Variable Selection}\label{simusel}

To evaluate the methods' performance in terms of variable selection, we construct the Receiver Operating Characteristic (ROC) by varying selection thresholds $\lambda$ (for ORS, ORF, and \texttt{ordinalNet}) and calculate the Area Under this Curve (AUC) in each iteration. Briefly, we evaluate the estimated models as a binary classification task, where covariates are predicted to be present or absent.  
For \texttt{polr}, the order of variables entering the (forward) stepwise selected model is used for ranking. Variables that were not selected at all are randomly coerced to the back.
We only discuss the results of scenario (a) from Figure~\ref{hoshiyar:fig6a} in detail here, where true effects contain unfused effects of $5$ levels. Due to a lack of space, we refer to the online supplementary material for the other settings' results. 

\begin{figure}[!ht]\centering
\includegraphics[width=12cm]{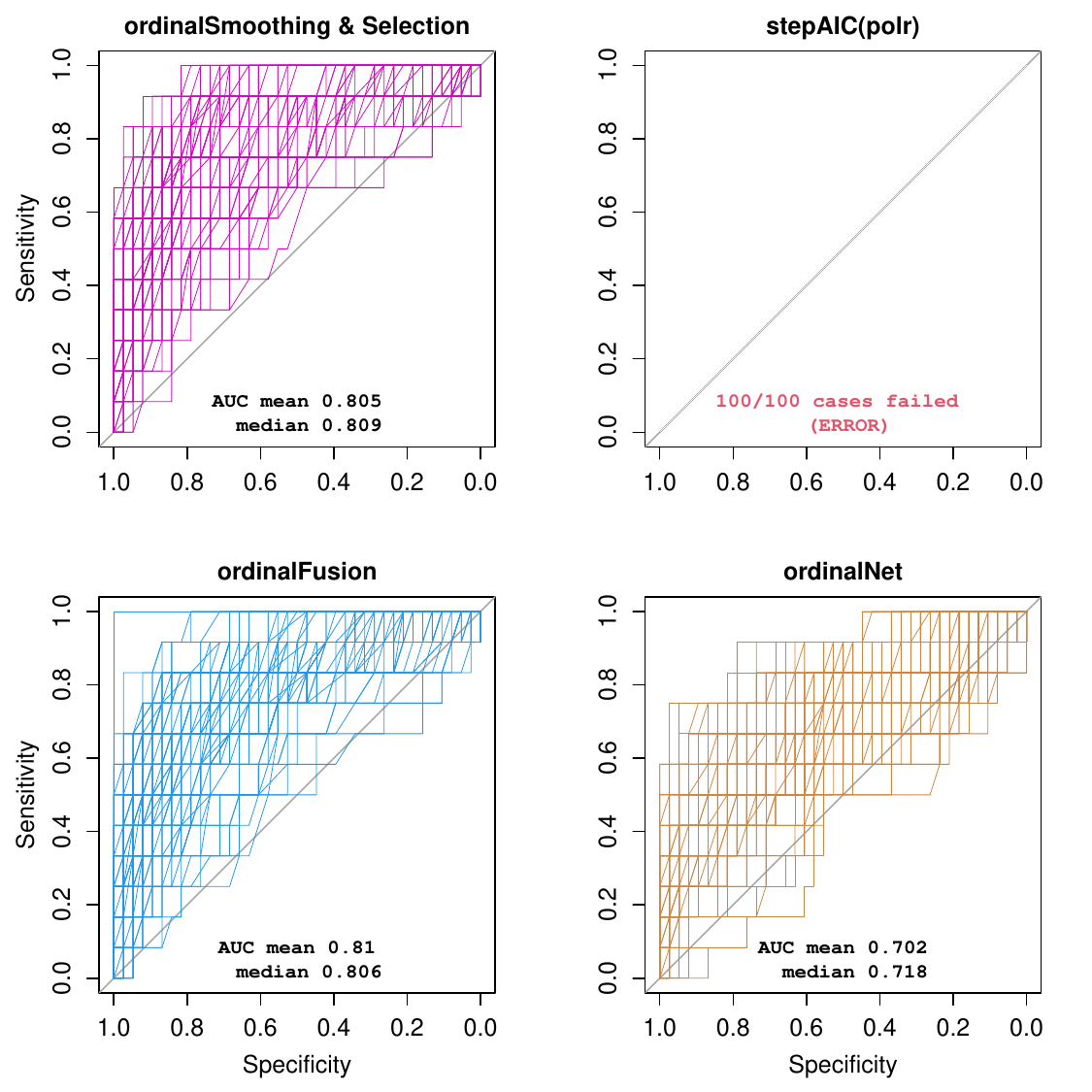} 
\captionof{figure}{\label{hoshiyar:fig6b} ROC curves when using ORS, \texttt{polr}, ORF, or \texttt{ordinalNet} with $n=200$; according to simulation setting (a) from Figure~\ref{hoshiyar:fig6a} ($5$ levels without fused effects).}
\end{figure} 

\begin{figure}[!ht]\centering
\includegraphics[width=0.85\linewidth]{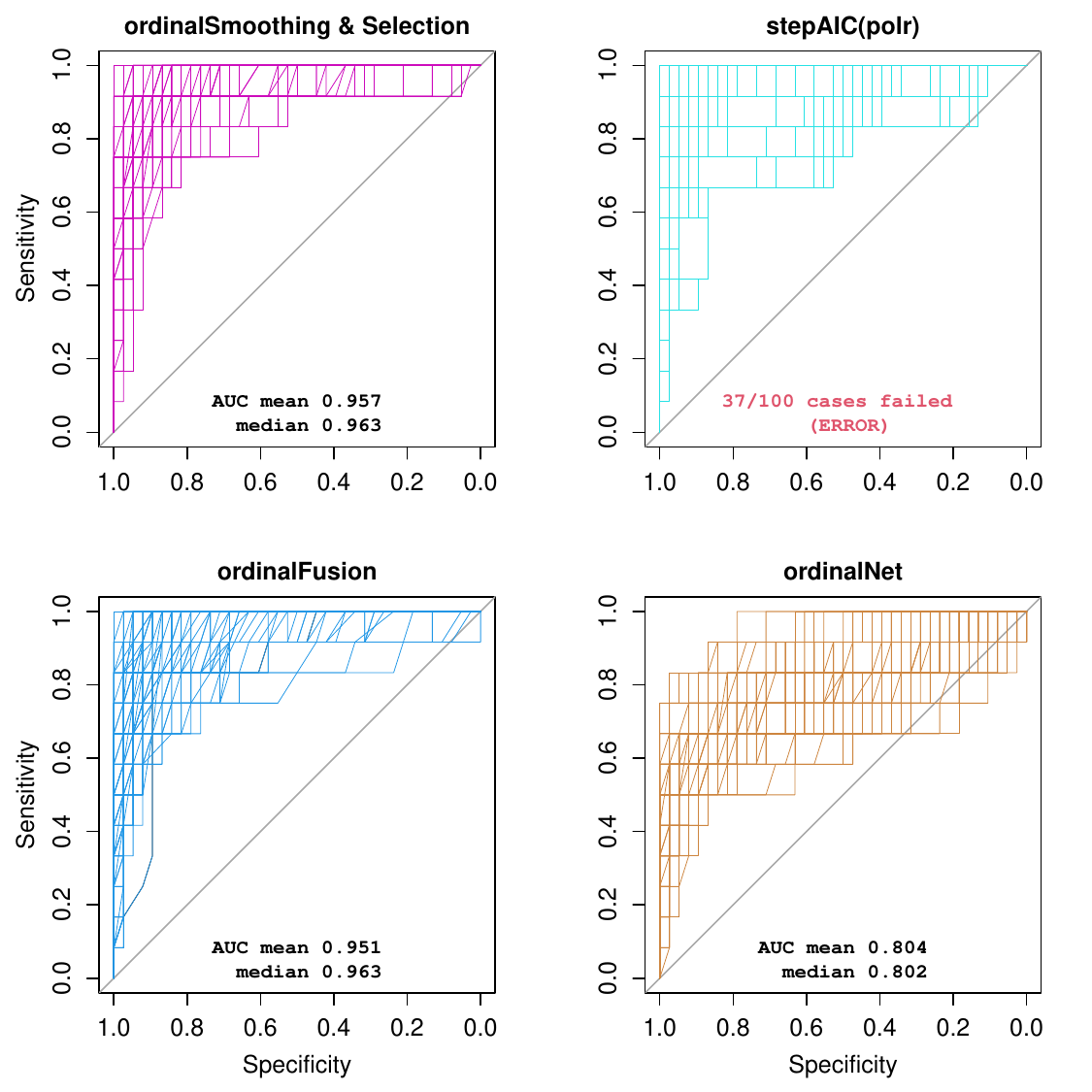} 
\captionof{figure}{\label{hoshiyar:fig6c} ROC curves when using ORS, \texttt{polr}, ORF, or \texttt{ordinalNet} with $n=500$; according to simulation setting (a) from Figure~\ref{hoshiyar:fig6a} ($5$ levels without fused effects).}
\end{figure} 

Figures~\ref{hoshiyar:fig6b} and \ref{hoshiyar:fig6c} show the performance in terms of the ROC as obtained with varying $\lambda$ values, $n=200$ and $n=500$, respectively. Additionally, the mean and median AUC for each method is given. The results for the remaining scenarios can be found in the online supplementary material (Figures~C1--C12). 
Even with the relatively large sample size of $500$, \texttt{polr} failed in $37$ of the datasets (zero cases with $n=1000$, but all cases with $n=200$). This result is in line with the issues that we observed with the standard POM for the boar-taint data in Section~\ref{ioi}. The mean and median AUC over the 63 successful runs with $n=500$ were $0.908$ and $0.919$, respectively. By contrast, the ordinal penalties work very well, and both account for the ordinal-on-ordinal nature. 
In general, the choice between penalties (\ref{defJ}) and (\ref{fus}) depends on the concrete application and personal preferences.
While both enforce the selection of predictors, ORS yields smooth effects of predictors, whereas ORF yields parameter estimates that tend to be flat over some neighboring categories, which represents a specific form of smoothness, too. The classical \texttt{ordinalNet} approach, however, is inferior for model selection since it offers no group lasso option for categorical/ordinal predictors.
With $n=1000$ (not shown here), all methods --except for classical lasso-- work (almost) equally well in scenarios~(a) and (c). 
In scenarios (b) and (d), i.e., with predictors having nine levels each, \texttt{polr} performs even worse than Figure~\ref{hoshiyar:fig6c}~(top right), as estimation now fails in all cases with $n=500$. Even with $n=1000$, \texttt{polr} is inferior to penalized selection and fusion in the nine predictor level setting (compare the online appendix for details, Figures~C4--6 and C10--12).

To gain more insights into the different behaviors of the competing methods, we tested whether differences in AUC values are statistically significant. Table~C2 in the online appendix presents the p-values of the corresponding (paired) t-tests. It is seen that the performance of the proposed penalties ORS and ORF does not significantly differ in most cases. Graphical comparisons of AUC values confirm these findings (Figures~C13 and C14 in the online appendix). 
It is a positive result that the ROC and AUC of the considered fusion and smoothing/selection penalty are very similar, indicating that the fusion penalty performs well for variable selection purposes, too.


\subsubsection{Level Fusion}

The results discussed so far concern the identification of relevant predictors. In scenarios (c) and (d), however, we are also interested in comparing the performance regarding the identification of relevant differences (i.e., fusion between adjacent dummy coefficients.  
For this purpose, we consider estimates using penalty (\ref{fus}). 
\begin{figure}[!ht]\centering
\includegraphics[width=0.85\linewidth]{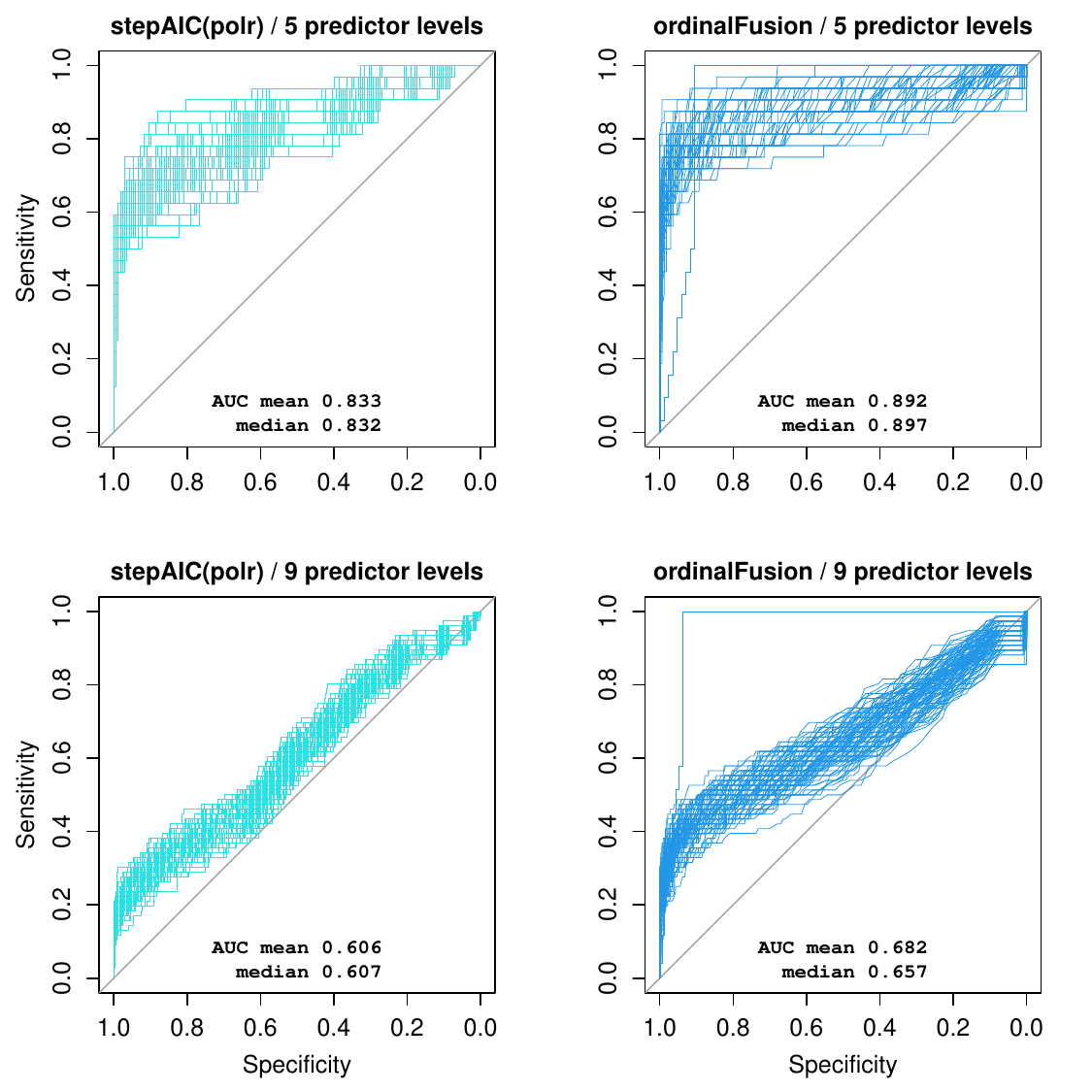} 
\captionof{figure}{\label{hoshiyar:fig6e} ROC curves when using \texttt{polr} or \texttt{ordinalFusion} concerning the identification of relevant differences (i.e., fusion) of dummy coefficients for $n=1000$. Top: $5$ levels with fused effects; bottom: $9$ levels with fused effects; i.e., simulation setting (c) and (d) from Figure~\ref{hoshiyar:fig6a}, respectively.}
\end{figure} 
Figure~\ref{hoshiyar:fig6e}~(right panel) shows the performance in terms of ROC as obtained with varying $\lambda$ and $n=1000$. In addition, we run \texttt{stepAIC(polr)}~(Figure~\ref{hoshiyar:fig6e}, left panel) as a competing model, but on the re-coded design matrix~\citep{Walter:1987}, thus selecting (presumably relevant) differences between neighboring categories. 
The run time of \texttt{polr} increases dramatically as contrasts of factor levels now enter and leave the model in each step separately. 
%
%
Figure~C15 in the supplementary material illustrates the false positive/negative rates (FPR/FNR).
With five underlying levels (scenario (c)) and $n=500$, \texttt{stepAIC(polr)} only converged in 23 cases, and, not surprisingly, in zero cases for $n=200$. This is why we only present results for the largest sample size $n=1000$ in Figure~\ref{hoshiyar:fig6e}. With nine underlying levels (scenario (d)), estimation with \texttt{polr} failed in all cases with $n=200$ and $n=500$. 
The poorer performance compared to variable selection (see also Figures~C9 and~C12, online appendix) can be explained by the fact that with selection, it is sufficient for a correct result (in terms of a `true positive') if at least one contrast was selected. For identification of the relevant differences, however, all contrasts of neighboring categories are checked individually. With predictors having nine levels each (Figure~\ref{hoshiyar:fig6e}, bottom row), the model has even more parameters that have to be taken into account (almost twice as many parameters as with five levels), which further worsens the results compared to the top row of Figure~\ref{hoshiyar:fig6e}.  
However, the fusion penalty still works better than \texttt{polr} with stepwise selection.


\section{Application to Case Studies}\label{sec5}

\subsection{Perception and Acceptance of Boar-tainted Meat}

\begin{figure}[!ht]\centering
 \includegraphics[width=\linewidth]{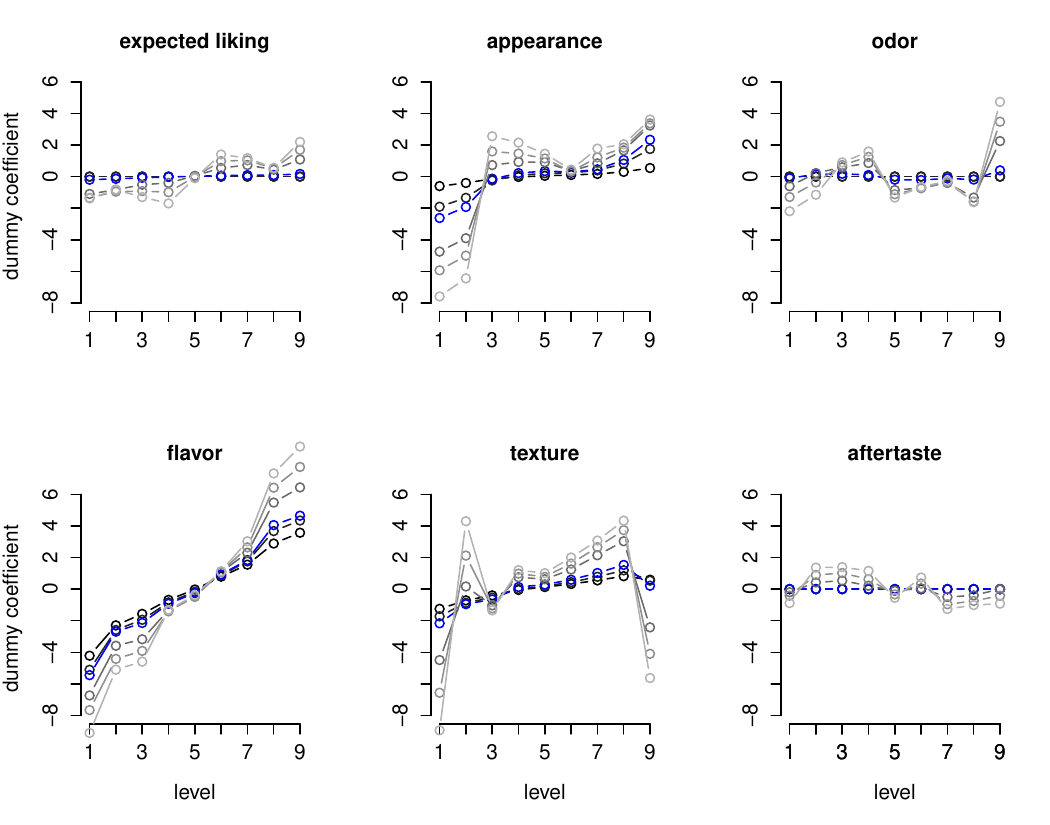} 
\caption{\label{hoshiyar:fig1} 
Cumulative group lasso estimates of the regression coefficients as functions of class labels ($\lambda \in \{10,5,1,0.5,0.25\}/n$) for the sensory data from Section~\ref{boart}. Light gray indicates smaller $\lambda$. Blue lines correspond to cross-validated/optimal $\lambda = 3.5/n$.}
\end{figure}

To illustrate the application of penalized ordinal-on-ordinal regression in practice, we first consider the boar-taint example from Section~\ref{boart}. The study aimed to identify important factors for consumer acceptance of boar-tainted meat. Determining the most important predictors of overall liking would be of great interest for the development of palatable products. Figure~\ref{hoshiyar:fig1} shows the resulting parameter estimates for the six covariates and different values of tuning parameter $\lambda$ when applying the proposed groupwise selection and smoothing method. It is seen that for smaller $\lambda$ (light gray), the estimates are more wiggly (and more extreme) and become more and more smoothed out/shrunken as $\lambda$ increases. It is also seen that, at some point, the entire group of dummy coefficients belonging to one variable is set to zero, which means that the corresponding variable is excluded from the model.  
\begin{figure}[!ht]\centering
 \includegraphics[width=\linewidth]{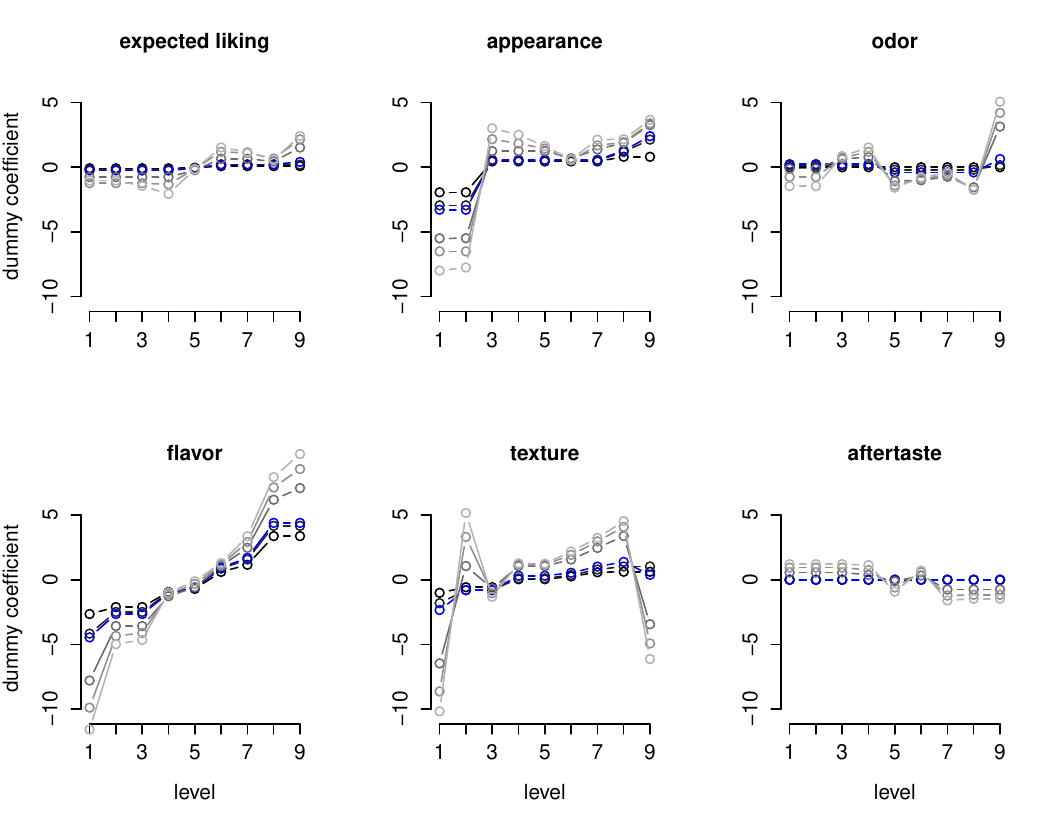} 
\caption{\label{hoshiyar:fig4} 
Cumulative fused lasso estimates of the regression coefficients as functions of class labels ($\lambda \in \{10,5,1,0.5,0.25\}/n$) for the sensory data from Section~\ref{boart}. Light gray indicates smaller $\lambda$. Blue lines correspond to cross-validated/optimal $\lambda = 4/n$.}
\end{figure}
Figure~\ref{hoshiyar:fig4} shows the fitted coefficients for the covariates from Figure~\ref{hoshiyar:fig1} when the fusion penalty~(\ref{fus}) is employed.
It is seen that some of the categories are fused, indicating that, at least for some variables, fewer rating categories would be sufficient to model the effect on overall liking. In general, the overall liking increases with the liking level of the covariates, with flavor apparently having the largest effect, followed by appearance and texture. On the other hand, the expected liking, the odor, and the aftertaste seem to be much less relevant. The obvious consequence of these findings is that, to make meat from entire male pigs more attractive to consumers, the producers (of processed meat) should try to make their products look and, most importantly, taste better, e.g., using spices masking the boar taint~\citep{MoeEtal:2019}. 

\begin{figure}[!ht]\centering 
\includegraphics[width=0.85\linewidth]{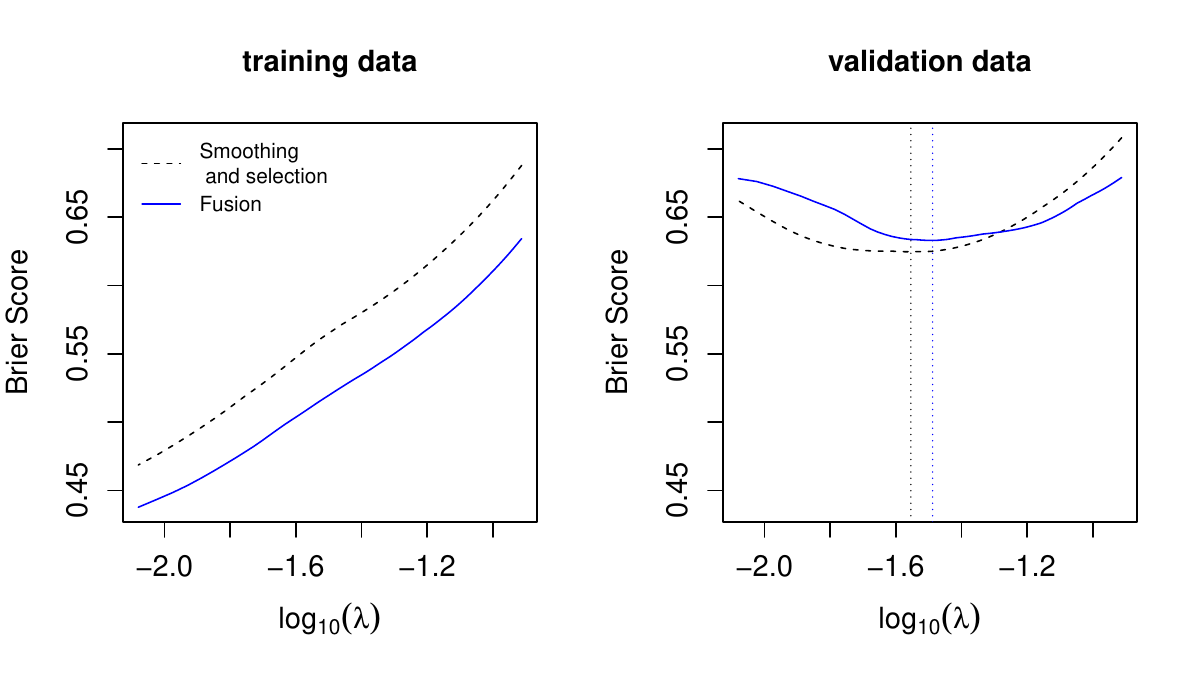}
\caption{\label{hoshiyar:fig5b} Brier Score (averaged over all folds) with ordinal smooth selection penalty (black dashed), and with ordinal fusion penalty (blue solid) for the sensory data from Section~\ref{boart}.
}
\end{figure}

The predictive performance according to 5-fold cross-validation on the boar-taint data for both the smooth group and fused lasso as a function of $\lambda$ is given in Figure~\ref{hoshiyar:fig5b}.
On the training data, this function is monotonically increasing in $\lambda$, as with smaller $\lambda$, more emphasis is put on the data. 
On the validation data, however, we see that the unpenalized proportional odds model (see $\lambda \to 0$, i.e., $\log_{10}(\lambda) \to - \infty$) is worse than smoothed selection or fusion. 
Cross-validation results thus indicate that predictive performance can be enhanced using the suggested penalized method(s). Performance improves on the validation data up to a certain $\lambda$ value and deteriorates from there. Note that the shift in the curves for the training data is not to be understood in the y-direction but in the x-direction. This is because the penalties are not directly comparable with each other.
The optimal smoothing parameter, as determined on the validation set(s), is indicated by the dotted lines in Figure~\ref{hoshiyar:fig5b}~(right). Based on those results, we can use the optimal $\lambda$ (where the Brier Score on the validation data reaches its minimum) to fit the final model. The resulting estimates are marked in blue in Figures~\ref{hoshiyar:fig1} and \ref{hoshiyar:fig4}, respectively.

\begin{figure}[!ht]\centering
 \includegraphics[width=0.85\linewidth]{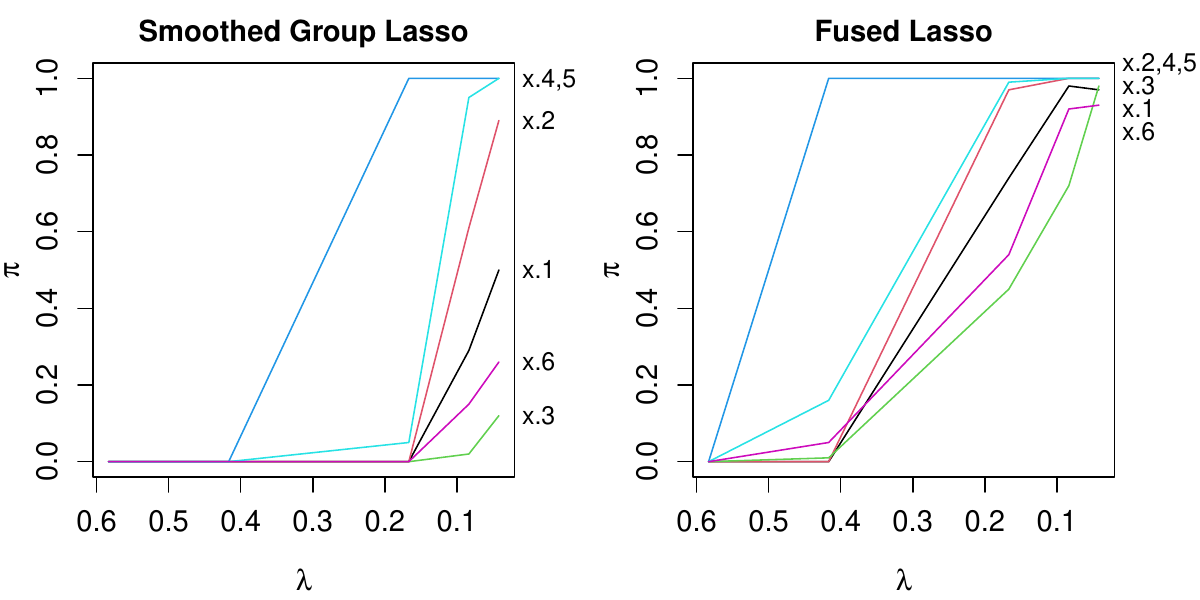} 
\caption{\label{hoshiyar:fig3} 
Stability path of the cumulative smooth/ordinal lasso (left) and the cumulative fused lasso (right) for the sensory data from Section~\ref{boart}.}
\end{figure}

Figure~\ref{hoshiyar:fig3} shows the stability paths for the selection penalty~(\ref{defJ}) and the fusion penalty~(\ref{fus}), indicating the order of relevance of the predictors according to stability selection. According to Figure~\ref{hoshiyar:fig3}, the probability to be selected within resampling is highest for {\it flavor}, {\it texture}, and {\it appearance} (in decreasing order), which confirms the interpretation from above. 
 
\subsection{Consumers' Willingness to Pay for Luxury Food}

We use penalties (\ref{defJ}) and (\ref{fus}) when modeling the luxury food data as described in the Introduction and Section~\ref{luxury}. The descriptive statistics on the data analyzed can be found in the online supplementary material, covering, among other things, a description of the items along with observed frequencies (Table~C3 and Figure~C16) and the correlation plot (Figure~C22). 
The code used for the presented case study is given in the vignette of our R package \texttt{ordPens} (\url{https://htmlpreview.github.io/?https://github.com/ahoshiyar/ordPens/blob/master/inst/ordinal-on-ordinal.html}). For more background information, we refer to \citet{Hartmann:Significance} who performed a partial least squares structural equation analysis (treating, however, the data as numeric). Using an initial principal component analysis (also treating the data as numeric), the authors revealed seven relevant dimensions concerning consumers' attitudes toward luxury food: hedonism, quality, sustainability, materialism, usability, uniqueness, and price. In our analysis, by contrast, the research question is more specific in that we try to identify factors that determine the consumers' willingness to pay for food products that they associate with luxury.

\subsubsection{Modeling Results}

\begin{figure}[tb]\centering
 \includegraphics[width=\linewidth]{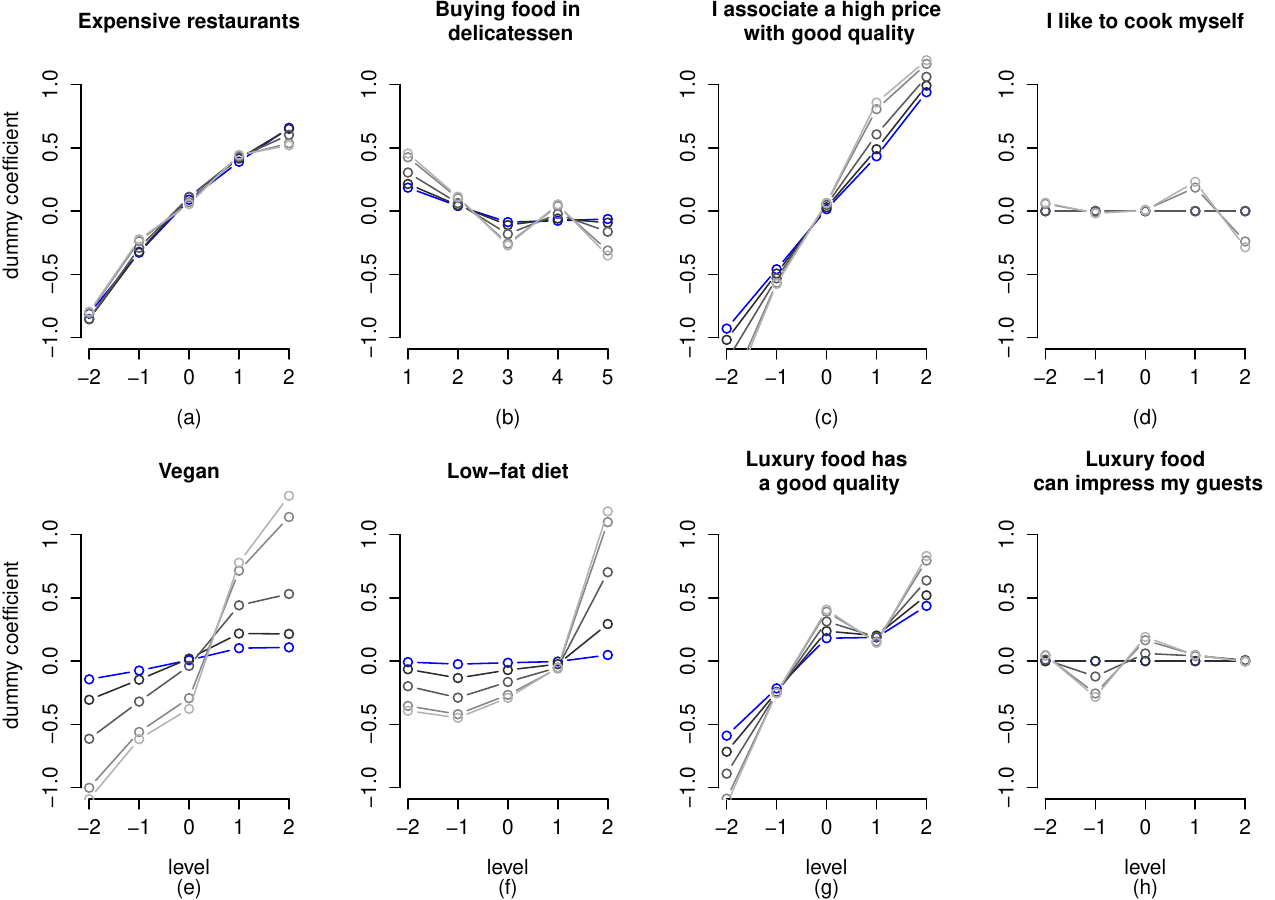} 
\caption{\label{hoshiyar:fig7a} 
Cumulative group lasso estimates of the regression coefficients as functions of class labels ($ \lambda \in {14.5,10,5,1,0.25}/n$) for the luxury food data from Section~\ref{luxury}.
Light gray indicates smaller $\lambda$.
}
\end{figure}
\begin{figure}[tb]\centering
 \includegraphics[width=\linewidth]{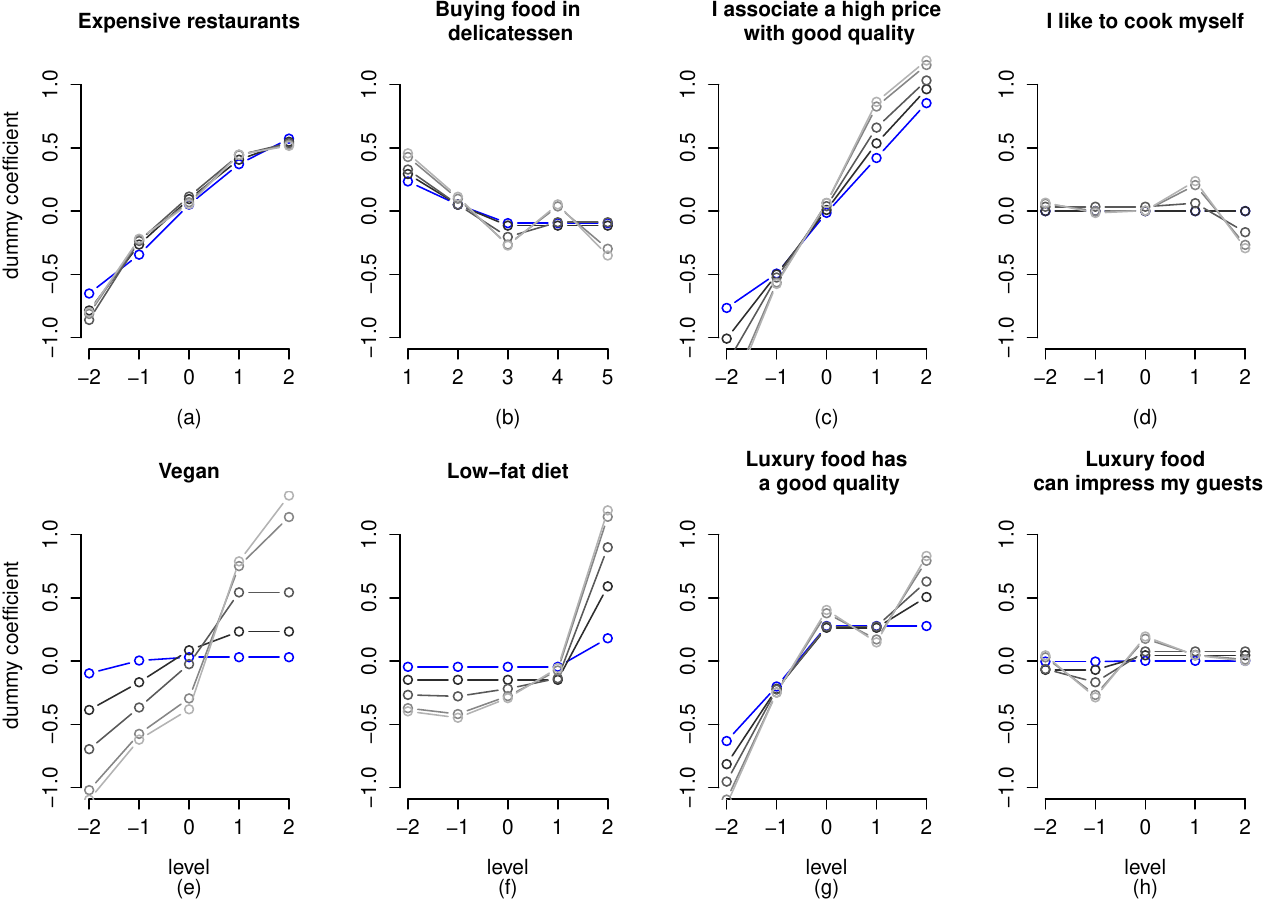} 
\caption{\label{hoshiyar:fig7b} 
Cumulative fused lasso estimates of the regression coefficients as functions of class labels ($ \lambda \in {18.5,10,5,1,0.25}/n$) for the luxury food data from Section~\ref{luxury}. Light gray indicates smaller $\lambda$.
}
\end{figure}

Figure~\ref{hoshiyar:fig7a} illustrates the resulting parameter estimates for selected covariates and different values of tuning parameter $\lambda$ when applying the ordinal group lasso. Figure~\ref{hoshiyar:fig7b} shows the corresponding results for the ordinal fused lasso. Similarly to the boar data example, the estimates are more wiggly for smaller $\lambda$ (light gray) but become smoother and shrunken as $\lambda$ increases. Especially, the willingness to pay tends to increase among people who (a) eat out frequently in expensive restaurants, (c) associate a high price with good quality, 
and (g) associate luxury with high quality. 
Our findings thus contribute to the literature on food and luxury by providing additional information about the kind of consumer who may be willing to pay a higher price for `luxury food'. For some items, e.g., ``I like to cook myself'' and ``Luxury food can impress my guests'', however, we observe rather flat effects that quickly go to zero as soon as $\lambda$ increases (compare the most right panel in Figure~\ref{hoshiyar:fig7b}), indicating that those items, and, hence, corresponding attitudes, provide only limited information regarding the respondent's willingness to pay for luxury food. As a consequence for, e.g., a producer of luxury food, it might not be worth targeting consumers (e.g., through a TV ad where a party host impresses the guests by serving luxury food) who primarily want to `show off' (compare Figure~\ref{hoshiyar:fig7b}, bottom right). Instead of that, the food producer should rather emphasize the food's high quality (compare Figure~\ref{hoshiyar:fig7b}, third panel). From a more general perspective, \citet{WiedmannEtal:2009} and \citet{Shirai:2015} already discussed the perceived association of luxury, quality, and price. \citet{AlfSha:2010} and \citet{ParsaEtal:2017} specifically considered customers in restaurants, highlighted price as a perceived positive indicator for the quality of food and investigated ways to increase the willingness to pay, respectively. An overview of the estimated effects of all items considered in our study is given in the supplementary material (Figures~C17--C21). Since the effects of penalties (\ref{defJ}) and (\ref{fus}) are quite similar, only the results of the fused lasso are presented there.

\begin{figure}[!t]\centering 
\includegraphics[width=0.85\linewidth]
{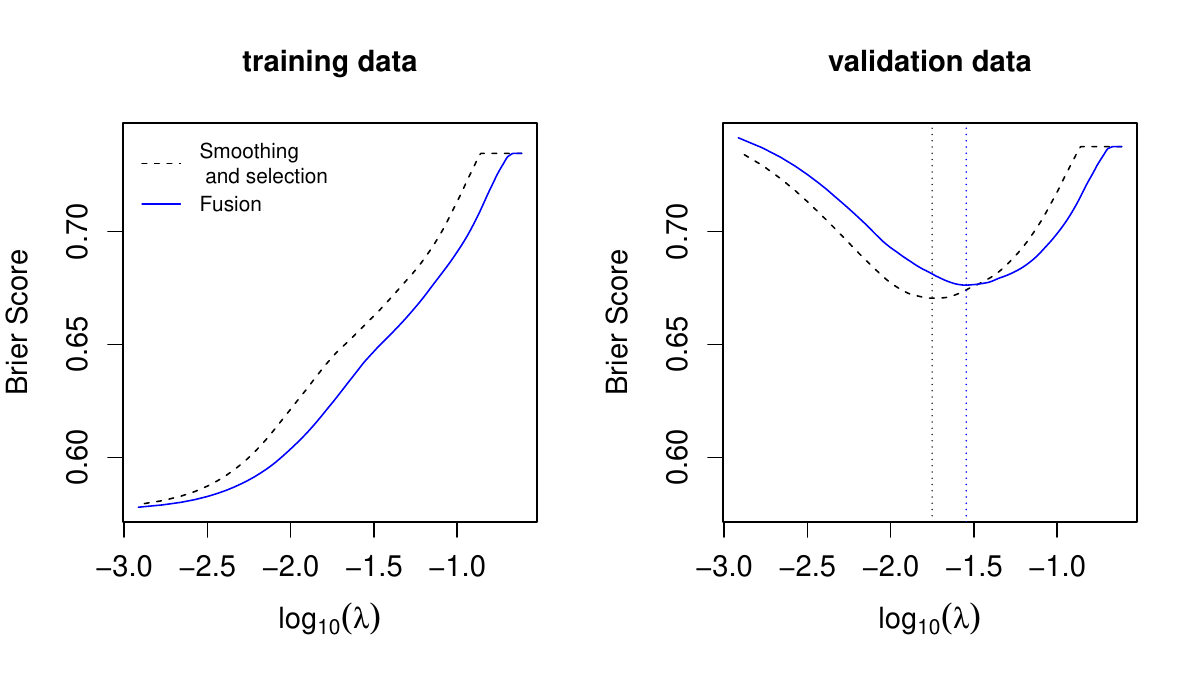} 
\caption{\label{hoshiyar:fig7c} Brier Score (averaged over all folds) with ordinal smooth selection penalty (black dashed),
and with ordinal fusion penalty (blue solid) for the luxury food data from Section~\ref{luxury}.
}
\end{figure}

The (5-fold) cross-validated predictive performance in terms of the Brier score as a function of $\lambda$ is given in Figure~\ref{hoshiyar:fig7c}.
If using the fused lasso and choosing the amount of penalty through (5-fold) cross-validation, we obtain an optimal $\lambda$ around $18.5/n$ (marked by the dotted blue line in Figure~\ref{hoshiyar:fig7c}, right). The corresponding estimates of the covariates' effects are highlighted in blue in Figure~\ref{hoshiyar:fig7b}. For the variable (f) ``Luxury food has a good quality'', for example, it is seen that the upper three categories are fused. If using the ordinal group lasso instead, the optimal (cross-validated) $\lambda$ is around $14.5/n$ (marked by the dotted black line in Figure~\ref{hoshiyar:fig7c}, right), with the resulting estimates in Figure~\ref{hoshiyar:fig7a} indicated by the blue lines.
If choosing $\lambda=14.5/n$ to obtain a final model, 17 variables are removed, and 26 variables are selected by the ordinal group lasso (with some showing very small effects, though). The most influential factors are those already shown in Figure~\ref{hoshiyar:fig7a}: (a) {\it expensive restaurants},  
(c) {\it associating a high price with good quality}, and (f) {\it associating luxury with high quality}. 
With the fused lasso, however, we might be more interested in detecting significant differences. To understand the fitted model's complexity, we refer to the degrees of freedom (df).  
For the fused lasso, an approximation is given by 
the number of non-zero parameters, i.e., non-zero differences of adjacent (dummy) coefficients
 \citep[cf.][]{Tibshirani:2005}. 
If using $\lambda=18.5/n$ for the ordinal fused lasso, we detect 51 relevant differences, which corresponds to about one-third compared to a total of $43 \times 4=172$ parameters in the full model. 

\subsubsection{Computational Complexity and Performance Evaluation}

With the fusion penalty, in total $3.9$ seconds elapsed when estimating the final model (i.e., using the optimal amount of penalization according to Figure~\ref{hoshiyar:fig7c}) on a MacBook with a 2.3 GHz processor and 16 GB RAM. With the smoothing/selection penalty, $5.7$ seconds elapsed. It should be noted, though, that the computation for a grid of $\lambda$ values does not increase the computational time by a factor of the length of the grid compared to a single $\lambda$. This is because our algorithm sorts the provided $\lambda$ in decreasing order and utilizes the results from the previous $\lambda$ value for a `warm start'. This approach significantly reduces the computational burden when working with multiple $\lambda$ values. For instance, running our ordinal group lasso algorithm (on the complete data set) on a grid of five different $\lambda$ values (as done to create Figure~\ref{hoshiyar:fig7a}) took approximately $14.9$ seconds.

  \begin{figure}[!ht]\centering
 \includegraphics[width=0.8\linewidth]{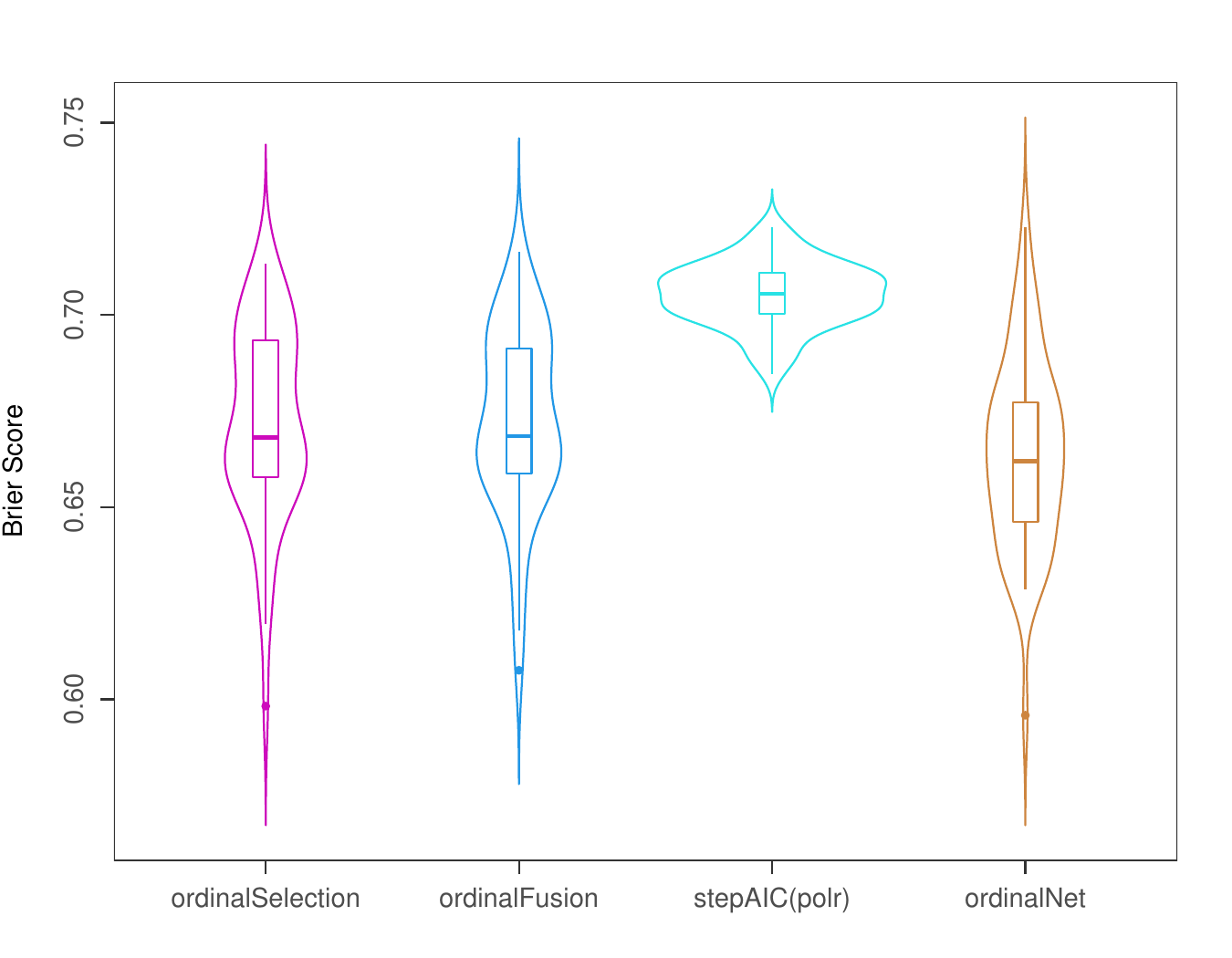} 
 \captionof{figure}{\label{hoshiyar:fig7d}Goodness of fit in terms of the test set Brier score (i.e., lower values are better) for $50$ random splits of the luxury food data from Section~\ref{luxury}. }
 \end{figure} 

In order to evaluate the goodness-of-fit of the models obtained using the proposed penalties~(\ref{defJ}) and (\ref{fus}), we performed repeated random splitting of the luxury data into training and test sets. That means all regression parameters are estimated on the training data (including the choice of the tuning parameter) and then used to predict the test data. As the test set size, we choose 100, and the procedure is independently repeated 50 times. As before in the simulations studies, we compare our ordinal group and fused lasso with the \texttt{ordinalNet} and a common proportional odds model/AIC-based stepwise selection. Resulting violin plots are shown in Figure~\ref{hoshiyar:fig7d}. Performance is again measured in terms of the test set Brier score. In the simulation studies in Section~\ref{simusel}, we saw that the common proportional odds model with factor variables performed well in large samples and was superior to \texttt{ordinalNet}. In the simulation studies, however, we had a controlled setting with partially non-linear and even non-monotonic effects. Results on the luxury data set are somewhat different. Although the data set has a moderate sample size, the unpenalized POM seems to attribute too much importance to random noise in the data and, therefore, tends to overfit. The considered penalty methods, on the other hand, perform equally well, indicating that the covariates' effects are approximated reasonably well by linear functions of the integer-valued predictor levels (as the latter is done by \texttt{ordinalNet}).

\section{Summary and Discussion}\label{sec6}

In the present work, we proposed two ways of regularization in ordinal-on-ordinal regression. By use of tailored penalty terms, these methods effectively extract information from ordinally scaled predictors and account for ordinal response variables. In particular, an $L_2$-difference penalty of first order for the selection of relevant factors and an $L_1$ variant for fusion of categories were presented. These approaches are more flexible than the methods existing so far for ordinal responses in combination with ordinal predictors, and allow for smoothing of covariate effects. The proposed methodology is implemented in R and publicly available on CRAN through add-on package \texttt{ordPens}.

To motivate and illustrate the proposed methods, and to show their potential benefits, 
we presented two real-world examples of item-on-items regression.
First, we considered a consumer test on the acceptance and perception of boar-tainted meat. The corresponding data set consisted of seven Likert-type items with nine levels each. 
In summary, our presented results suggest that penalized regression works well in the cumulative logit model with ordinally scaled predictors and ordinal group or fused lasso penalty. We combined the penalty approach with stability selection to gain additional information about a variable's probability to be included in the model, given a certain amount of penalization. Namely, if trying to predict the product's overall liking score, the stability path revealed that the probability to be selected (within resampling) is highest for the flavor, the texture, and the appearance of the meat product.

For comparison, we considered the standard proportional odds model implemented in the \texttt{polr()} R-function, which simply treats categorical/ordinal covariates as dummy-coded factors and offers no option for smoothing. 
We found that the use of \texttt{polr} can cause problems as (a) the employed fitting routine often fails (even for moderately dimensioned setups) and (b) often results in wiggly/implausible estimates of the (categorical) covariates' effects. Our numerical experiments distinguished between settings involving five or nine predictor levels. In summary, \texttt{polr} only worked well when the sample size was large enough and (often) failed otherwise. 
Ordinal selection and ordinal fusion worked equally well and outperformed \texttt{polr} in most scenarios. Overall, performance deteriorated with an increasing number of category levels, i.e., increasing model complexity.
In addition, we compared our smoothing/selection and fusion penalties to the classical lasso for the cumulative logit model as implemented in \texttt{ordinalNet()}. This method offers no group lasso option for the selection of categorical predictors. In our numerical experiments, it was thus not particularly surprising that in terms of AUC values, standard \texttt{ordinalNet} with linear effects performed worst and therefore seems somewhat unsuited for ordinal-on-ordinal regression if at least some of the effects may be non-linear across factor levels or even non-monotone. 
 
Our main interest was in a high-dimensional survey on luxury food expenditure, again consisting of Likert-type items. On the one hand, it can be assumed that in a data set of this size, not all of the (43) items will be relevant for explaining the response, making variable selection a primary concern. On the other hand, we would like to have estimates that are rather smooth and interpretable while taking the covariates' ordinal scale level into account. In summary, penalized (cumulative logit) regression appeared to be a sensible approach in studies of this type. 
For some items, we could observe that most categories were fused. For some (presumably) less relevant items, it could be seen that they reached effects equal to zero quickly as the amount of penalization increased.  
To obtain a final model,  an appropriate penalty parameter was chosen using cross-validation. With the group lasso, eventually, about 40\% of the covariates were removed from the model. 
With the fused lasso, we detected about one-third of the parameters to be relevant (in the sense of non-zero differences between adjacent categories). 

An alternative to the one-step, penalized likelihood methods presented in this paper would be to use regularization for model selection only (i.e., identification of the relevant covariates and relevant differences between predictor levels). Afterward, the final model would be re-fit on the reduced and collapsed data through common, unpenalized maximum likelihood. Such an approach was, e.g., used in \citet{GerTut:2010}, and is also known under the name `post-lasso'~\citep{BelEtAl:2012, BelEtAl:2013}. On the data sets considered in this article, however, we did not see any improvement in terms of out-of-sample prediction accuracy. Therefore, we did not present the results in detail here.


\section*{Acknowledgements}
A. Hoshiyar's and J. Gertheiss' research was supported by Deutsche Forschungsgemeinschaft (DFG) under Grant GE2353/2-1.


%
\singlespacing 

\beginsupplement

 \newgeometry{left=35mm, right=35mm, top=30mm, bottom=30mm}

%
\appendix
\section{Technical Details } 

In what follows, we present additional details and technical specifications related to the methods discussed in the main paper. Section~\ref{pom} gives the formulation of the cumulative logit model in more detail. Section~\ref{mle} describes the estimation process using Maximum Likelihood. Section~\ref{penalty} introduces the matrix that produces differences of first order used for the penalty, detailing the implementation of the ordinal fusion penalty.

\subsection{The Cumulative Logit Model}\label{pom}

Given the logistic distribution function $F(\eta_{ir})=1/(1+\exp(-\eta_{ir}))$, the cumulative logit model has the form
\begin{equation}\label{cumlogit}
P(y_i \leq r)=\frac{\exp(\theta_r + \mat{x}^\top_i \mat{\beta})}{1+\exp(\theta_r + \mat{x}^\top_i \mat{\beta})}.
\end{equation}
Let $(y_{i1},\ldots, y_{ic})$ be binary indicators of the response for subject $i$, with $y_{ir} = 1$ if $y_i=r$ and zero otherwise, and $\mat{\theta} = (\theta_1,\ldots,\theta_{c-1})^\top$ is the vector of thresholds. The log-likelihood for the sample is then given by
\begin{equation}\label{logl}
l(\mat{\theta}, \mat{\beta}) = \sum_{i=1}^n \sum_{r=1}^c y_{ir} \log{\pi_{ir}}, 
\end{equation}
with 
\begin{equation*}
\pi_{ir} = \left\{\begin{array}{ll} \frac{\exp{(\eta_{ir})}}{1+\exp{(\eta_{ir})}}, & r=1 \\
         \frac{\exp{(\eta_{ir})}}{1+\exp{(\eta_{ir})}} - \frac{\exp{(\eta_{i,r-1})}}{1+\exp{(\eta_{i,r-1})}}, & r=2,\ldots,c\end{array}\right. .
\end{equation*}

With ordinal covariates,
we introduce dummy variables $z_{ijl}$ and redefine the linear predictor as given in Equation~(2.1) of the main paper. The complete indicator design matrix is then given by
$\mat{Z} = (\mat{Z}_1|\ldots|\mat{Z}_p)$ and
$$\mat{Z}_j = (z_{ijl}) =
\begin{pmatrix}
z_{1j1} & \cdots & z_{1jk_j}\\
 \vdots &  \ddots & \vdots \\
 z_{nj1} & \cdots & z_{njk_j} \\
\end{pmatrix}.
$$

\subsection{Maximum Likelihood Estimation in the Cumulative Logit Model}\label{mle}

Estimation of unknown coefficients and thresholds $\theta_r$ relies on maximum likelihood principles that are described below. 
To embed the cumulative logit model into the framework of generalized linear models (GLMs), we first introduce the response function as well as the design matrix.
The occurrence probability $P(y_i=r)=P(y_{ir}=1)=\pi_{ir}$ in model~\eqref{logl} 
 is connected to the linear predictor $\eta_{ir}$ via the response function $h_r(\mat{\eta}_{i})$ 
as follows:
\newline
$\pi_{i1} = h_1(\eta_{i1},\ldots,\eta_{ic}) = F(\eta_{i1})$ and
$$
\pi_{ir} = h_r(\eta_{i1},\ldots,\eta_{ic}) = F(\eta_{ir})-F(\eta_{i,r-1}), \quad r=2,\ldots,c
$$
with
$$
\eta_{ir}=\theta_r + x_{i1} \beta_1 + \ldots + x_{ip} \beta_p
$$
and $F(\eta_{ir})=1/(1+\exp(-\eta_{ir}))$ being the logistic distribution function.

In matrix notation, we can write $\mat{\eta}_i = \mat{X}_i \mat{\gamma}$ with
\begin{equation*}
\mat{X}_i = 
\begin{pmatrix}
1 & & & & \mat{x}^\top_i \\ 
  & 1 & & & \mat{x}^\top_i \\ 
 & & \ddots & &  \vdots \\ 
 & & & 1 & \mat{x}^\top_i \\  
\end{pmatrix}
\end{equation*}
and the parameter vector $\mat{\gamma} = (\theta_1, \ldots, \theta_{c-1}, \mat{\beta}^\top)^\top$.

The $c$-dimensional response function is given by $\mat{\pi}_i = \mat{h}(\mat{\eta}_i)$.
Again let $(y_{i1},\ldots, y_{ic})$ be binary indicators of the response for subject $i$. Then, the log-likelihood for the sample has the form 
\begin{equation*}  
l(\mat{\gamma}) = \sum_{i=1}^n \sum_{r=1}^c y_{ir} \log{\pi_{ir}} =   \sum_{i=1}^n  \mat{y}_i \log \big\{ \mat{h}(\mat{\eta}_i)  \big\}.
\end{equation*} 
with 
\begin{equation*}
h_r(\eta_{ir}) = \left\{\begin{array}{ll} \frac{\exp{(\eta_{ir})}}{1+\exp{(\eta_{ir})}}, & r=1 \\
         \frac{\exp{(\eta_{ir})}}{1+\exp{(\eta_{ir})}} - \frac{\exp{(\eta_{i,r-1})}}{1+\exp{(\eta_{i,r-1})}}, & r=2,\cdots,c\end{array}\right. . 
\end{equation*} 
Based on the exponential family, the score function $\mat{s}(\mat{\gamma})=\partial l(\mat{\gamma}) / \partial \mat{\gamma} $ takes the form   
$$
\mat{s}(\mat{\gamma})= \sum_{i=1}^n \mat{X}_i^\top \mat{D}_i(\mat{\gamma}) \mat{\Sigma}_i^{-1}(\mat{\gamma}) [\mat{y}_i - \mat{h}(\mat{\eta}_i)], 
$$
where $\mat{D}_i(\mat{\gamma})=\partial \mat{h}(\mat{\eta}_i)/ \partial \mat{\eta}$ denotes the derivative of $\mat{h}(\mat{\eta})$ evaluated at $\mat{\eta}_i = \mat{X}_i \mat{\gamma}$ and $\mat{\Sigma}_i(\mat{\gamma})= \text{cov}(\mat{y}_i)$ is the covariance matrix of indicator observation vector $\mat{y}_i=(y_{i1},\ldots,y_{ic})$ given parameter vector $\mat{\gamma}$. By introducing the weight matrix $\mat{W}_i(\mat{\gamma}) = \mat{D}_i(\mat{\gamma}) \mat{\Sigma}_i^{-1}(\mat{\gamma}) \mat{D}_i^\top(\mat{\gamma})$, we can write the expected Fisher matrix as 
$$
\mat{F}(\mat{\gamma}) = \text{cov}(\mat{s}(\mat{\gamma})) = \sum_{i=1}^n \mat{X}_i^\top \mat{W}_i(\mat{\gamma}) \mat{X}_i = \text{E}(-\nabla^2 l(\mat{\gamma})).
$$
Note that in order to incorporate categorical/ordinal predictors, we have to replace $p$-dimensional vectors $\mat{x}_i^\top$ by corresponding dummy vectors $\mat{z}_{i} = (\mat{z}_{i1}^\top,\ldots,\mat{z}_{ip}^\top)^\top$, $\mat{z}_{ij}^\top = (z_{ij1},\ldots,z_{ijk_j})$. 
Then, $\mat{\gamma}$ 
collects all regression parameters (thresholds plus regression coefficients) to be estimated, potentially after reparametrization to respect constraints as described in Subsection~\ref{pom} and 2.2. 

\subsection{Penalty Matrices and Parameter Transformations}\label{penalty}

For the implementation of the smoothing and selection penalty given in Equation~(3.2) in the main paper, consider the vector $\mat{\beta}_j^\top = (\beta_{j1},...,\beta_{jk_j})$ for predictor $j$. The vector of first differences is given by $\tilde{\mat{\beta}}_j^\top = (\beta_{j2} - \beta_{j1} ,\beta_{j3} - \beta_{j2},...)$. It can be obtained as
$\tilde{\mat{\beta}}_j = \mat{D}_{1,j}\mat{\beta}_j$ with the $((k_j-1)\times k_j)$-dimensional difference matrix
\begin{equation}\label{defD}
\mat{D}_{1,j} =
\begin{pmatrix}
-1 & 1  & 0 & \cdots & 0 \\
0 & -1 & 1  & \cdots & 0 \\
0 & \ddots & \ddots & \ddots & 0 \\
0 & \cdots & 0 & -1 & 1 \\
\end{pmatrix}.
\end{equation} 
Then, the parameter estimates are computed as described in the main paper. 

If the fusion penalty given in Equation~(3.3) is used instead, we make use of \texttt{ordinalNet:ordinalNet()} \citep{Wurm:2021}, a function that provides ordinal regression models with elastic net penalty. If the elastic mixing parameter $\alpha \in [0,1]$ is set to \texttt{alpha=1}, the usual $L_1$ lasso penalty is applied. 
By appropriate reparameterization and recoding as proposed by \citet{Walter:1987}, the problem can be written as a problem in which the classical lasso penalty is used. 
The general split-coding scheme is as follows.
The transformed design matrix becomes $\mat{\tilde{X}}=(\tilde{\mat{X}_1}|\ldots|\tilde{\mat{X}}_p)$, $\tilde{\mat{\beta}}=(\tilde{\mat{\beta}}_1^\top,\ldots,\tilde{\mat{\beta}}_p^\top)^\top$ and $\tilde{\mat{X}}_j=\mat{X}_j\mat{D}^{-1}_{1,j}$, $\tilde{\mat{\beta}}_j=\mat{D}_{1,j}\mat{\beta}_j$. The new parameters now have again the form $\tilde{\mat{\beta}}_j=(\tilde{\beta}_{j,1},\ldots,\tilde{\beta}_{j,k_j-1})^\top$ with components $\tilde{\beta}_{j,l}=\beta_{j,l+1}-\beta_{j,l}$.
The split coding scheme is given in Table~\ref{Hoshiyar:tab21}. 

\begin{table}[!ht]\centering
\begin{tabular}{ccccccc}
  &\multicolumn{6}{c}{Dummy variable} \\
\cmidrule{2-7} %
Category & $z_{ij1}$ & $z_{ij2}$ & $z_{ij3}$ & & $z_{ij,k_j-2}$ & $z_{ij,k_j-1}$ \\
\midrule
1 & 0 & 0 & 0 & \ldots & 0 & 0 \\
2 & 1 & 0 & 0 & \ldots & 0 & 0 \\
3 & 1 & 1 & 0 & \ldots & 0 & 0 \\
\vdots & \vdots & \vdots & \vdots & & \vdots & \vdots \\
$k_j-1$ & 1 & 1 & 1 & \ldots & 1 & 0 \\
$k_j$ & 1 & 1 & 1 & \ldots & 1 & 1 \\
\end{tabular}\caption{\label{Hoshiyar:tab21} Values of the independent variables in each category following the split coding scheme.}
\end{table}

Note, that \texttt{ordinalNet} follows the common convention of scaling the negative log-likelihood by the number of observations, i.e.,
\begin{equation*}  
\begin{split}
l_\text{ordN}(\mat{\theta},\mat{\beta}) &= - \frac{1}{n} l(\mat{\theta},\mat{\beta}) + \lambda \sum_{j=1}^p J_j(\mat{\beta}_j)  \\
&= \frac{1}{n} \Bigl(- l(\mat{\theta},\mat{\beta}) + n \lambda \sum_{j=1}^p J_j(\mat{\beta}_j) \Bigr).
\end{split}
\end{equation*}

\newpage
\newgeometry{left=35mm, right=35mm, top=20mm, bottom=20mm}

\section{Detailed Results of the Simulation Studies}
 


\begin{figure}[H] 
\centering
\includegraphics[width=0.71\linewidth]{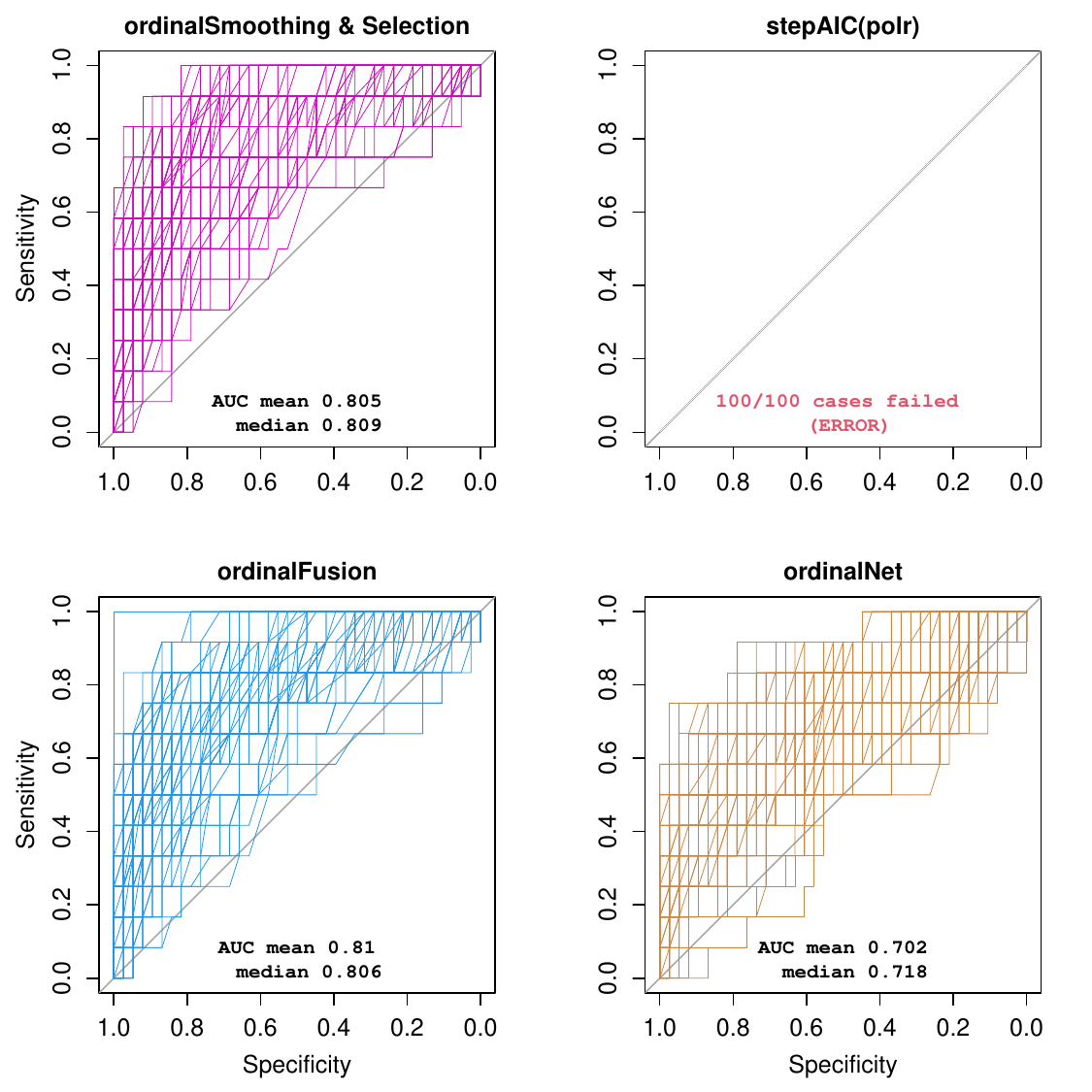}
\caption{\label{Figure_S1_1} 
ROC curves when using ORS, polr, ORF or ordinalNet with n = 200. According to setting (a) from Figure~3 (\textbf{$\mathbf{5}$ levels without fused effects}).  
}
\end{figure}

\begin{figure}[H] 
\centering
\includegraphics[width=0.71\linewidth]{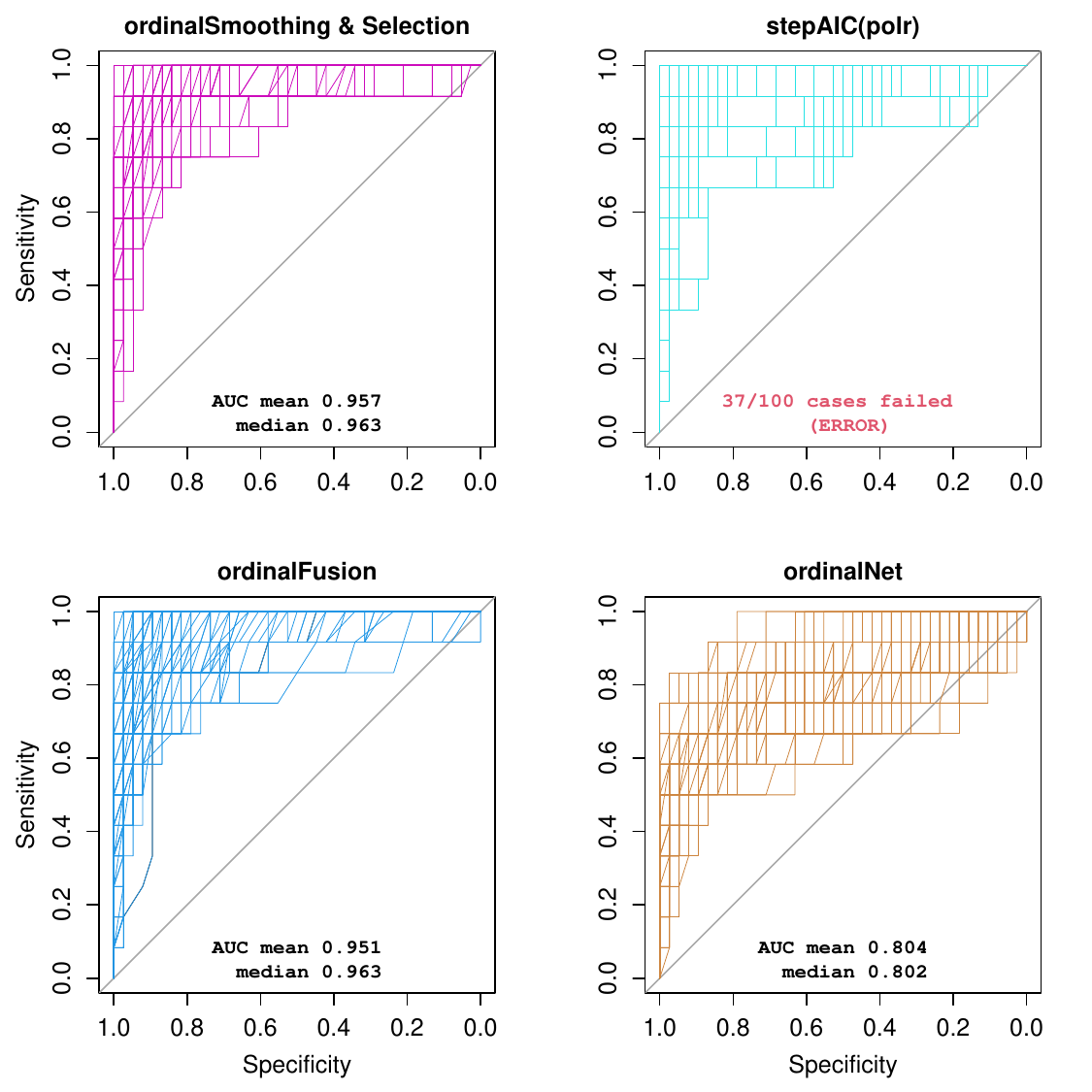}
\caption{\label{Figure_S1_2} 
ROC curves when using ORS, polr, ORF or ordinalNet with n = 500. According to setting (a) from Figure~3 (\textbf{$\mathbf{5}$ levels without fused effects}).  
}
\end{figure}

\begin{figure}[H] 
\centering
\includegraphics[width=0.71\linewidth]{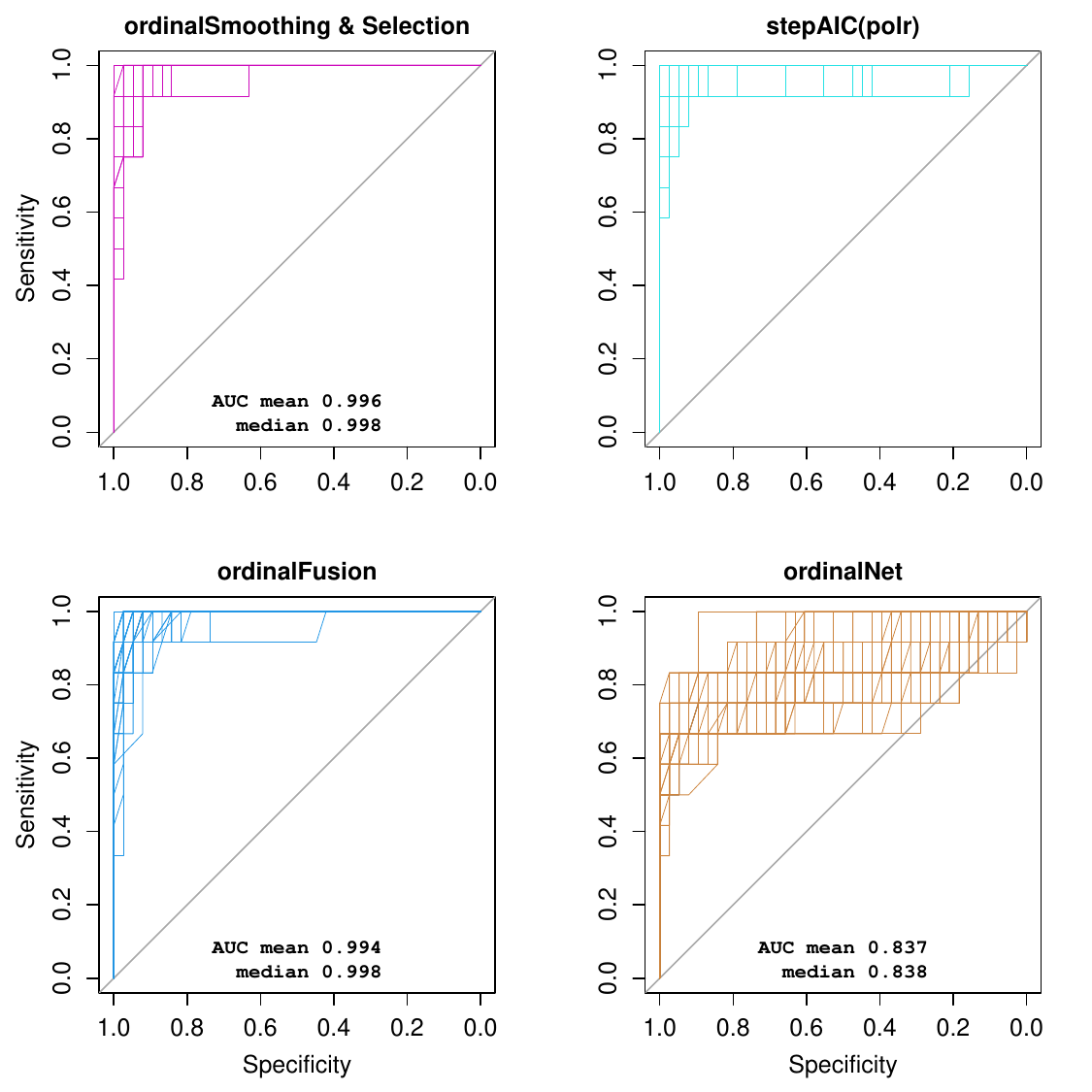}
\caption{\label{Figure_S1_3} 
ROC curves when using ORS, polr, ORF or ordinalNet with n = 1000. According to setting (a) from Figure~3 (\textbf{$\mathbf{5}$ levels without fused effects}).  
}
\end{figure}

\begin{figure}[H] 
\centering
\includegraphics[width=0.71\linewidth]{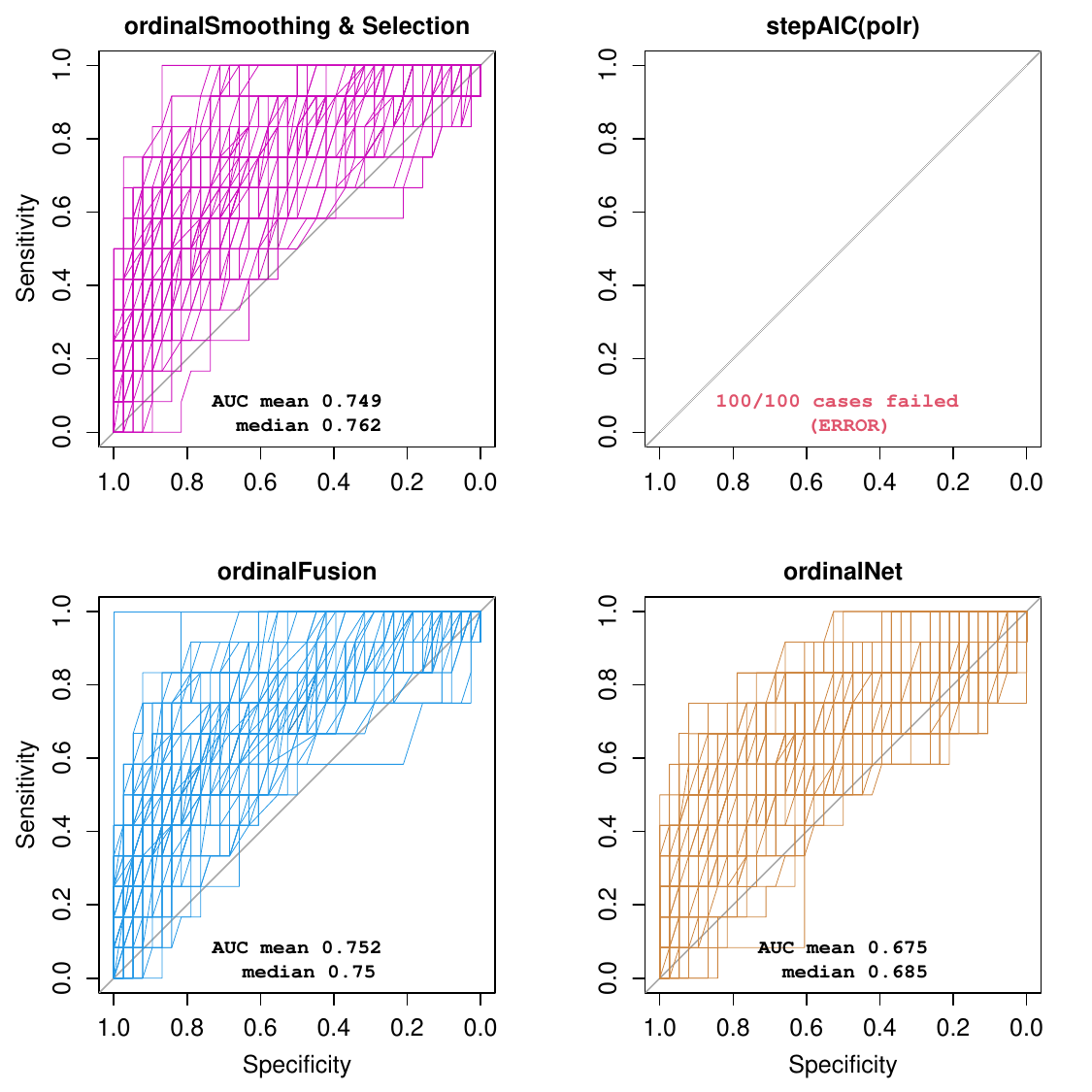}
\caption{\label{Figure_S2_1} 
ROC curves when using ORS, polr, ORF or ordinalNet with n = 200. According to setting (b) from Figure~3 (\textbf{$\mathbf{9}$ levels without fused effects}).  
}
\end{figure}

\begin{figure}[H] 
\centering
\includegraphics[width=0.71\linewidth]{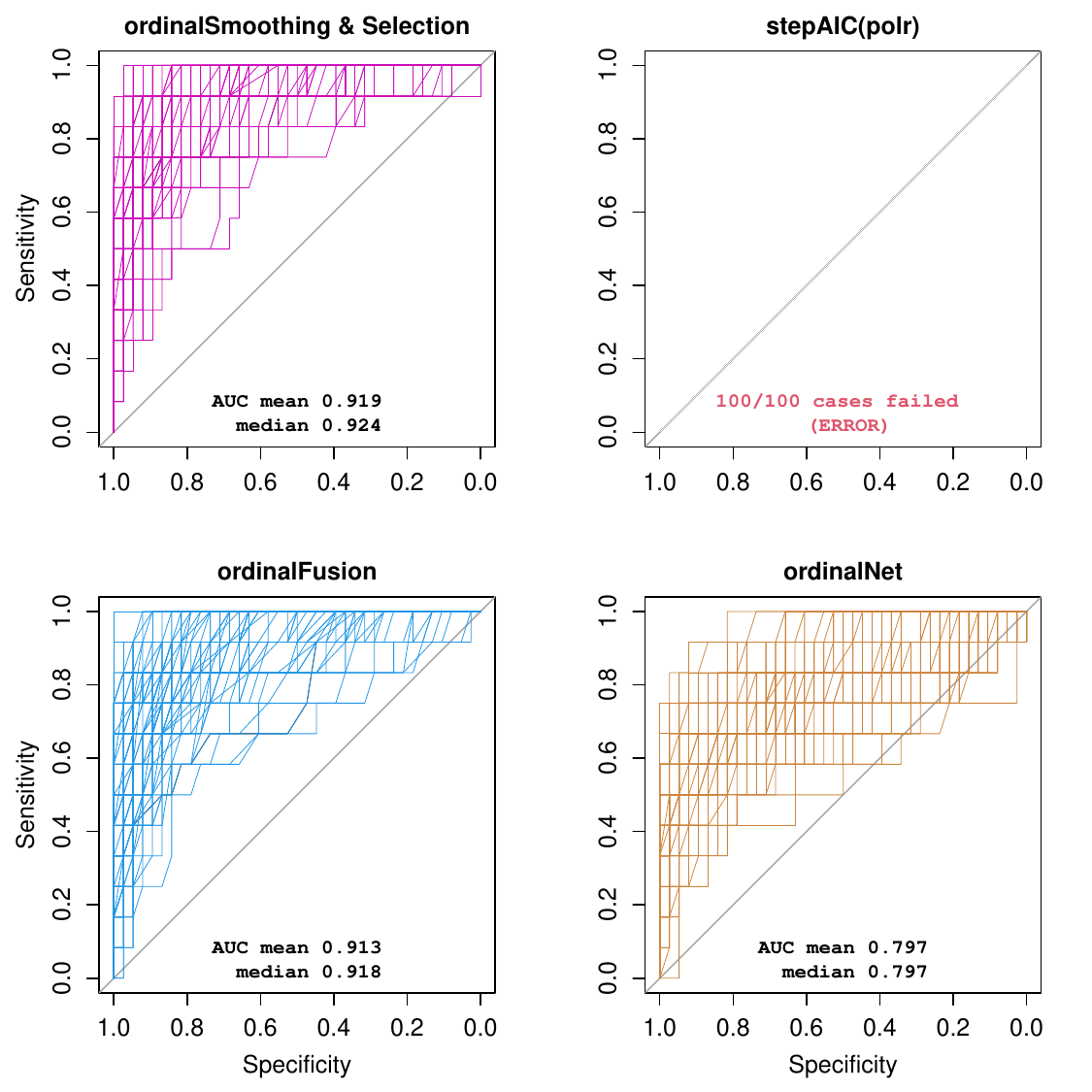}
\caption{\label{Figure_S2_2} 
ROC curves when using ORS, polr, ORF or ordinalNet with n = 500. According to setting (b) from Figure~3 (\textbf{$\mathbf{9}$ levels without fused effects}).  
}
\end{figure}

\begin{figure}[H] 
\centering
\includegraphics[width=0.71\linewidth]{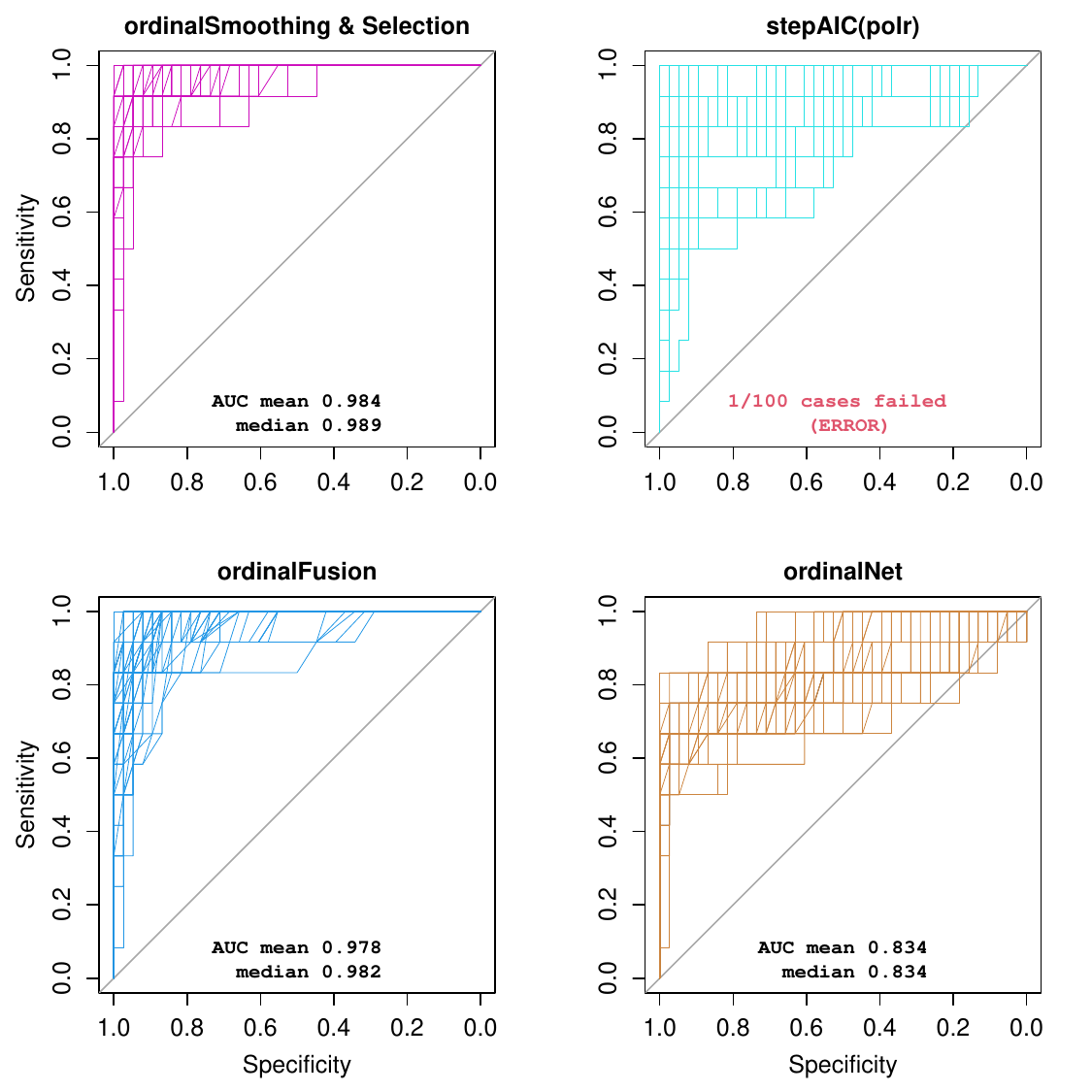}
\caption{\label{Figure_S2_3} 
ROC curves when using ORS, polr, ORF or ordinalNet with n = 1000. According to setting (b) from Figure~3 (\textbf{$\mathbf{9}$ levels without fused effects}).  
}
\end{figure}

\begin{figure}[H] 
\centering
\includegraphics[width=0.71\linewidth]{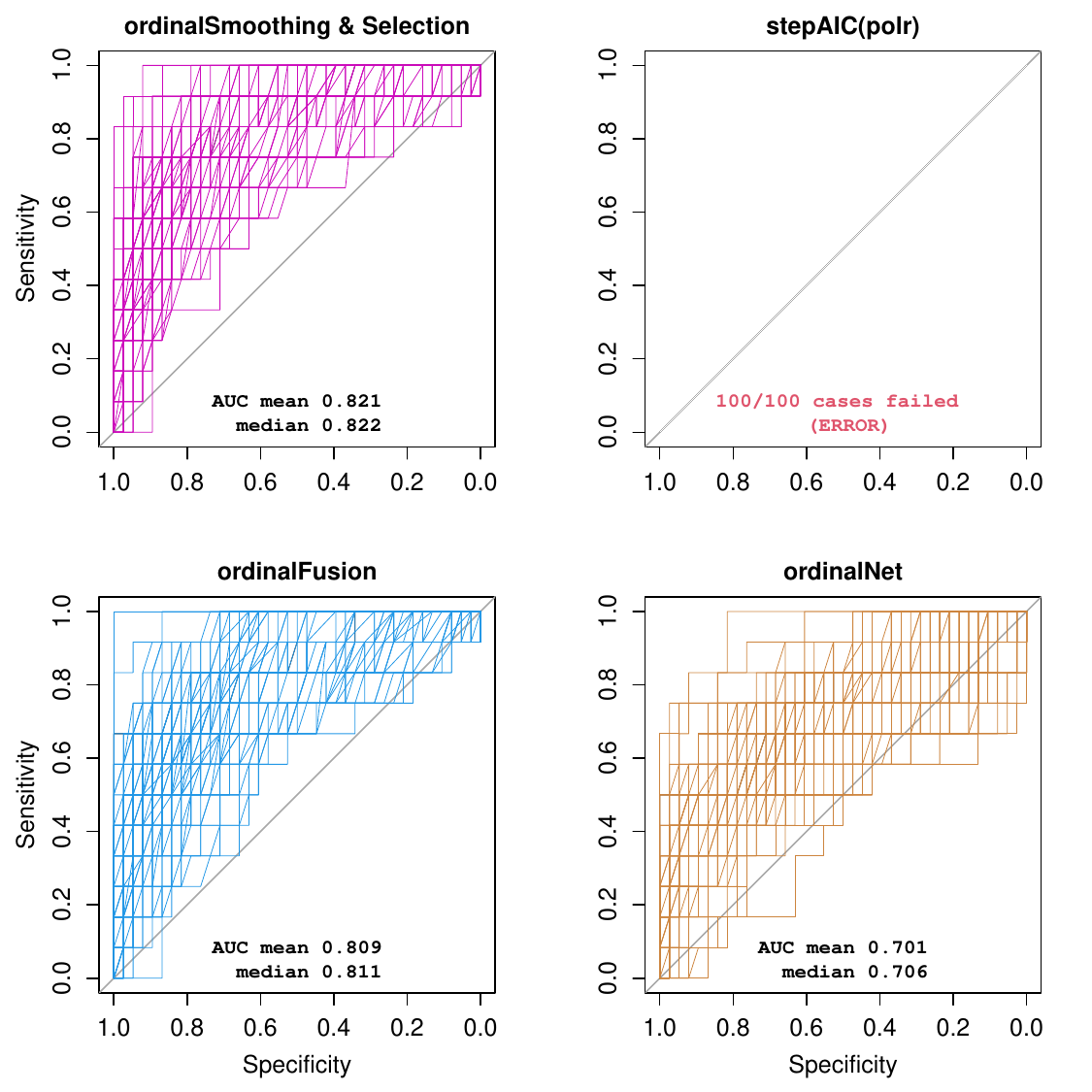}
\caption{\label{Figure_S3_1} 
ROC curves when using ORS, polr, ORF or ordinalNet with n = 200. According to setting (c) from Figure~3 (\textbf{$\mathbf{5}$ levels with fused effects}).  
}
\end{figure}

\begin{figure}[H] 
\centering
\includegraphics[width=0.71\linewidth]{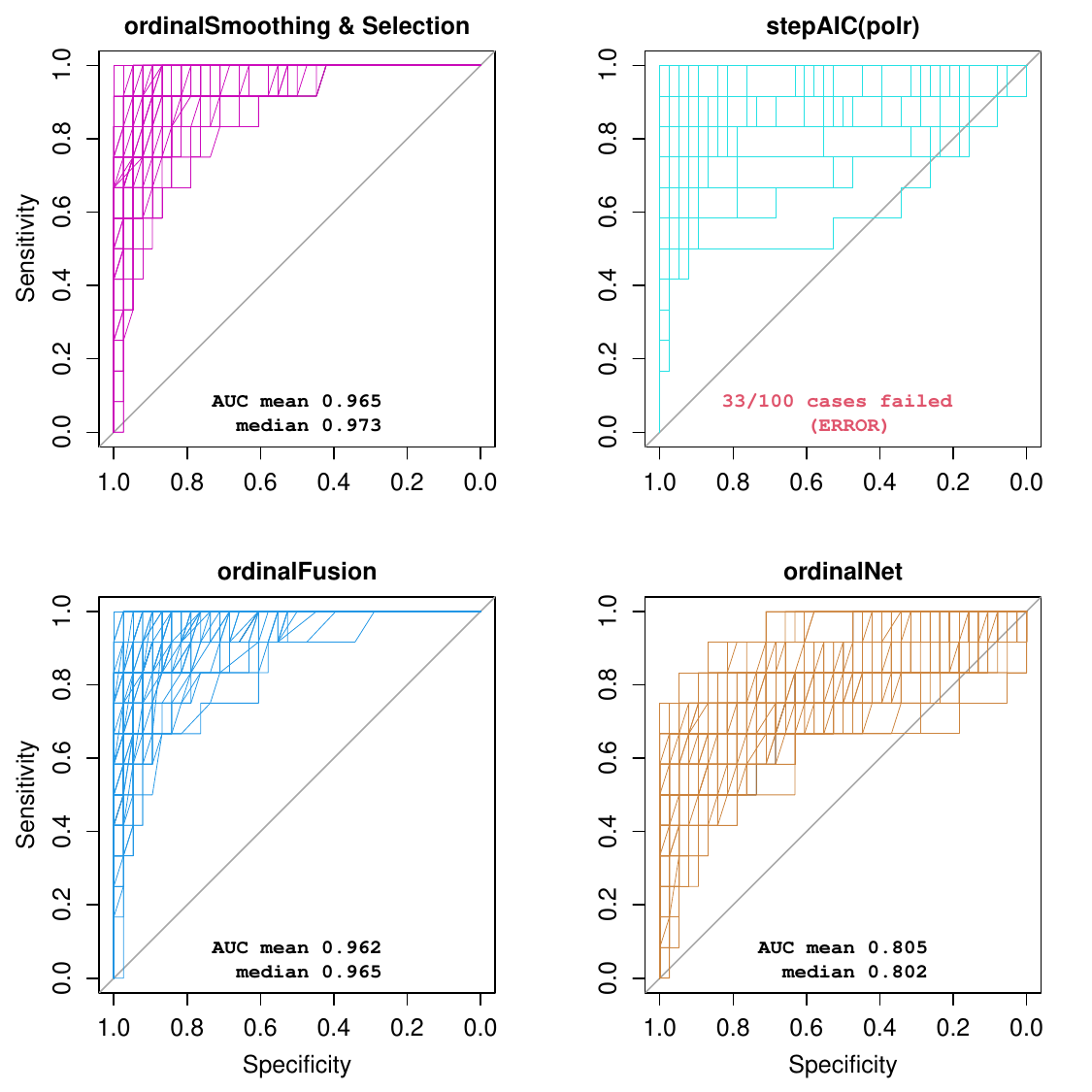}
\caption{\label{Figure_S3_2} 
ROC curves when using ORS, polr, ORF or ordinalNet with n = 500. According to setting (c) from Figure~3 (\textbf{$\mathbf{5}$ levels with fused effects}).  
}
\end{figure}

\begin{figure}[H] 
\centering
\includegraphics[width=0.71\linewidth]{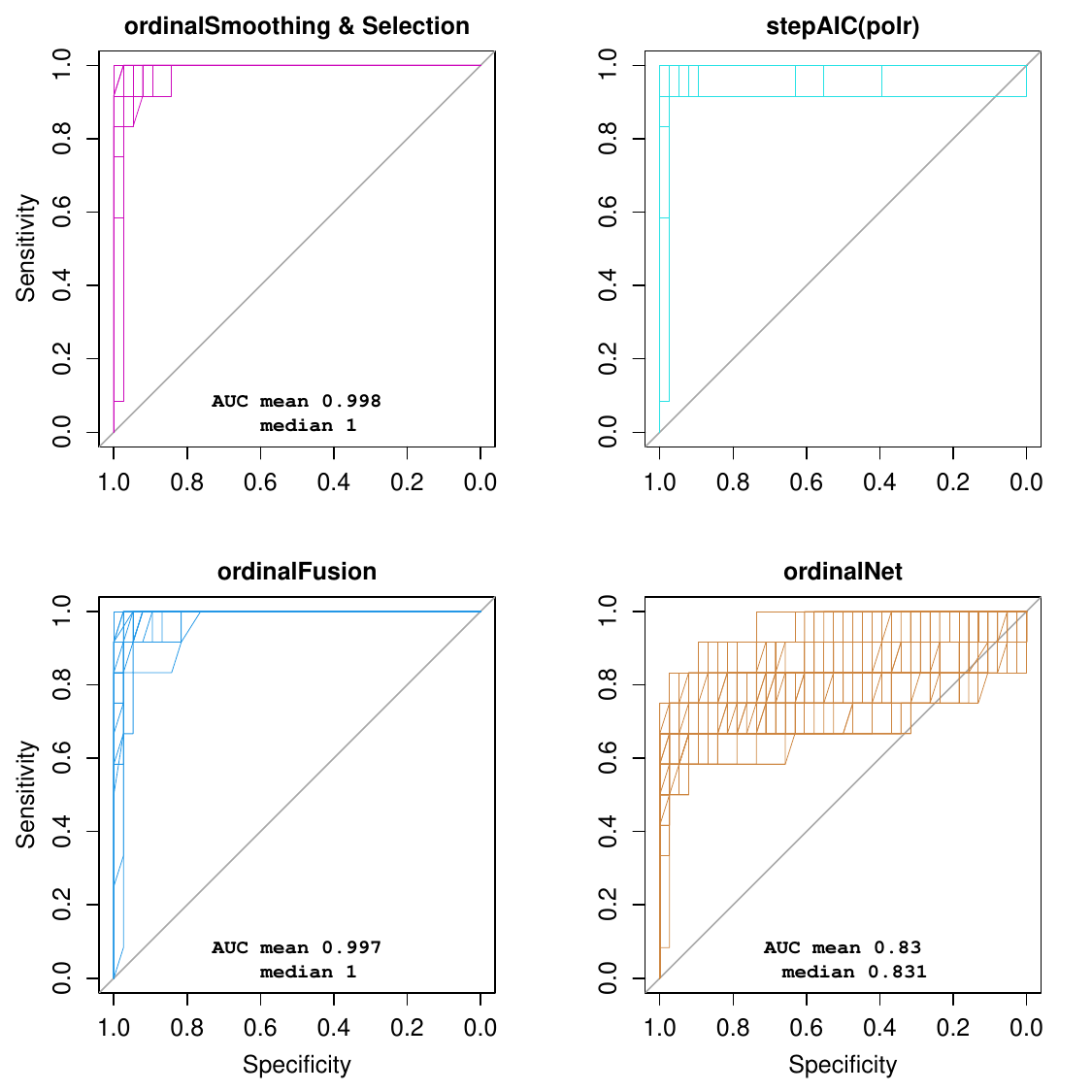}
\caption{\label{Figure_S3_3} 
ROC curves when using ORS, polr, ORF or ordinalNet with n = 1000. According to setting (c) from Figure~3 (\textbf{$\mathbf{5}$ levels with fused effects}).  
}
\end{figure}

\begin{figure}[H] 
\centering
\includegraphics[width=0.71\linewidth]{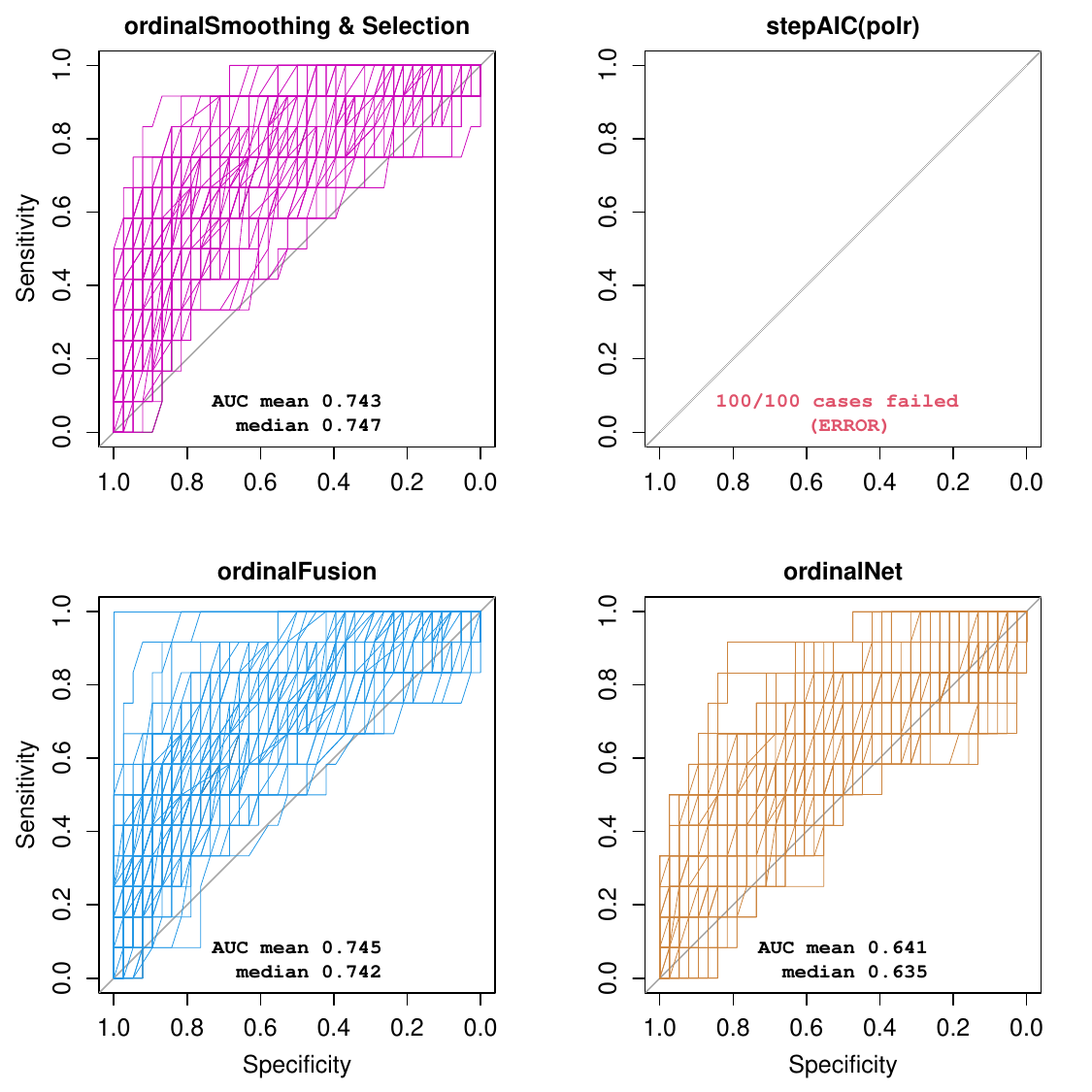}
\caption{\label{Figure_S4_1} 
ROC curves when using ORS, polr, ORF or ordinalNet with n = 200. According to setting (d) from Figure~3 (\textbf{$\mathbf{9}$ levels with fused effects}).  
}
\end{figure}

\begin{figure}[H] 
\centering
\includegraphics[width=0.71\linewidth]{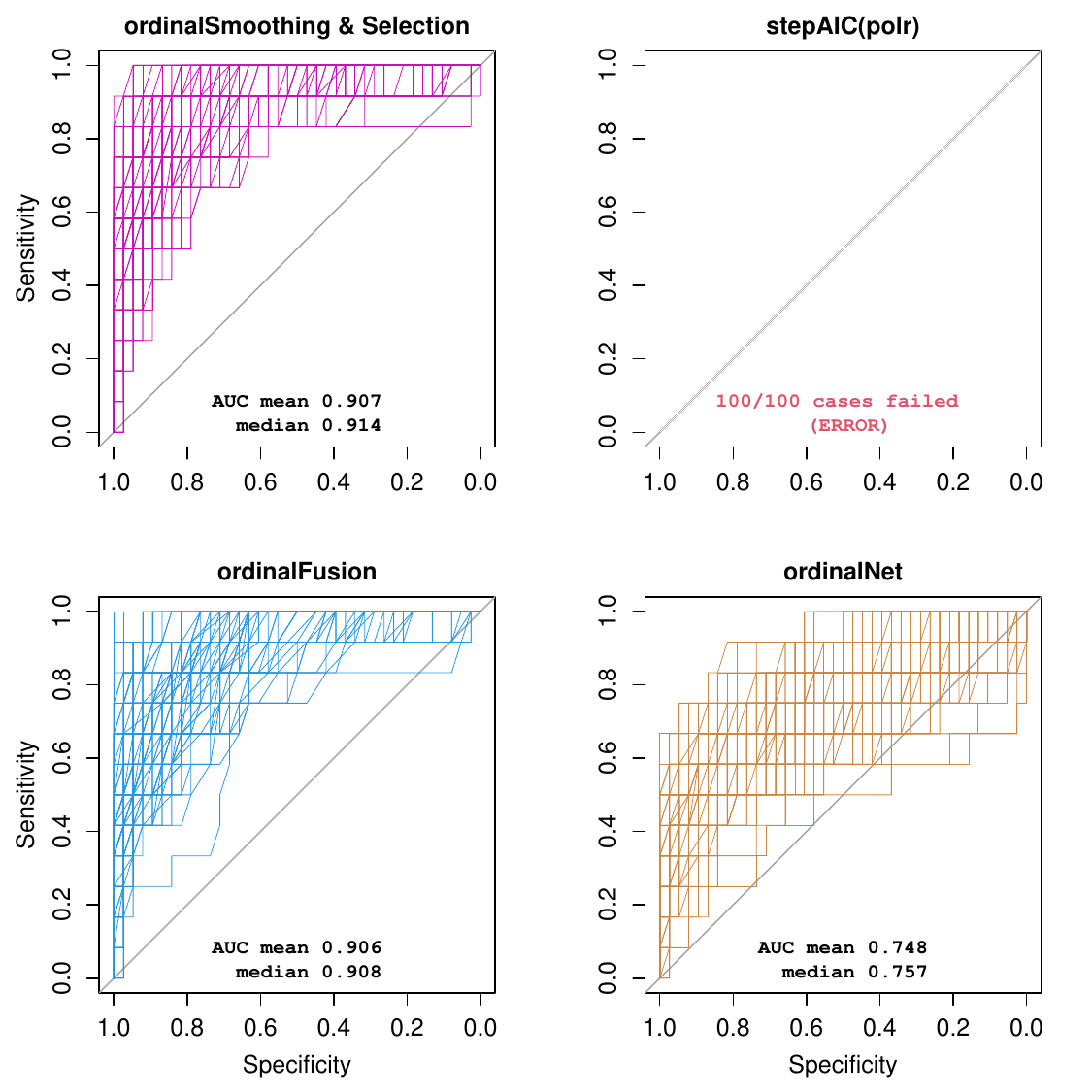}
\caption{\label{Figure_S4_2} 
ROC curves when using ORS, polr, ORF or ordinalNet with n = 500. According to setting (d) from Figure~3 (\textbf{$\mathbf{9}$ levels with fused effects}).  
}
\end{figure}

\begin{figure}[H] 
\centering
\includegraphics[width=0.71\linewidth]{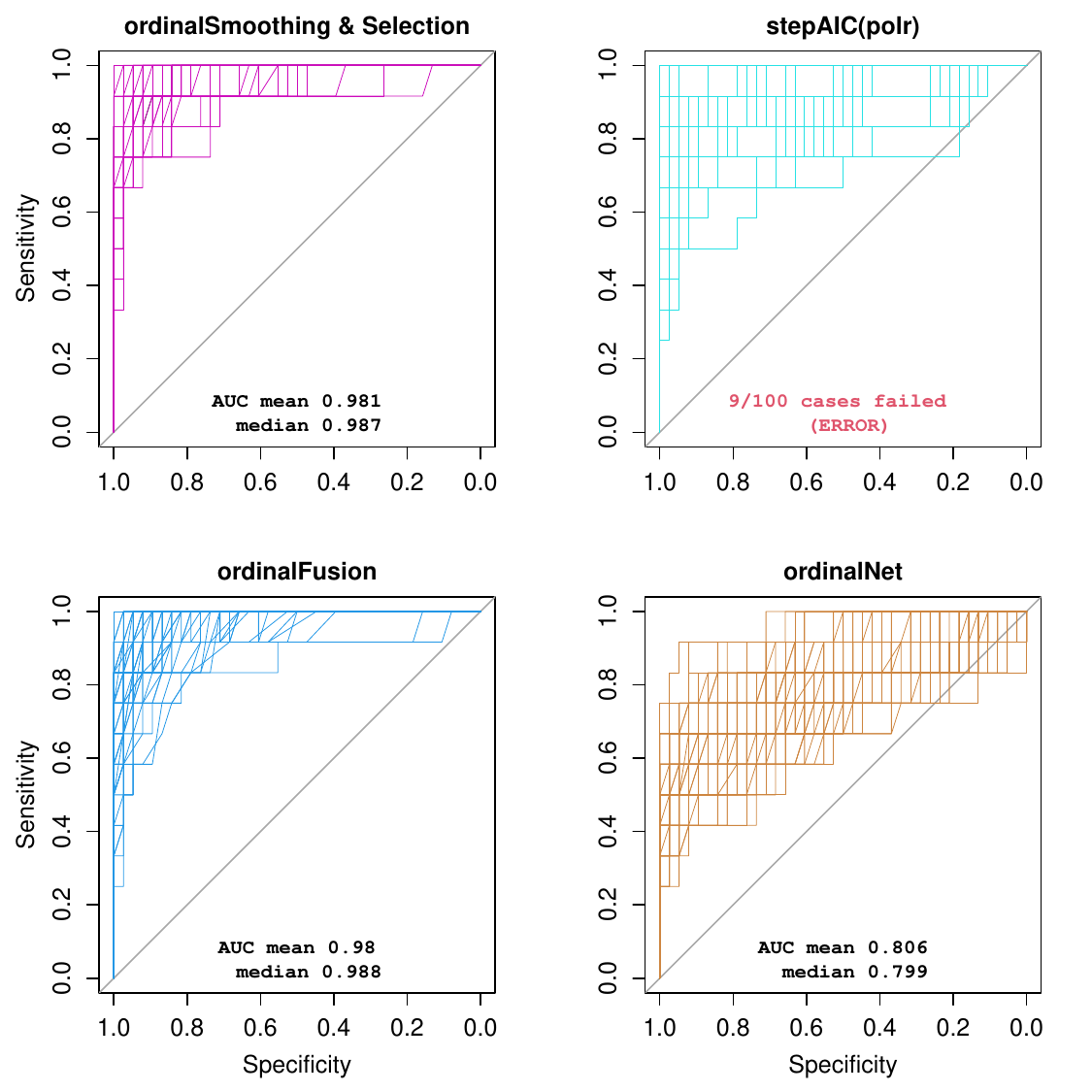} 
\caption{\label{Figure_S4_3} 
ROC curves when using ORS, polr, ORF or ordinalNet with n = 1000. According to setting (d) from Figure~3 (\textbf{$\mathbf{9}$ levels with fused effects}).  
}
\end{figure}

\begin{figure}[H]\centering
 \includegraphics[width=\linewidth]{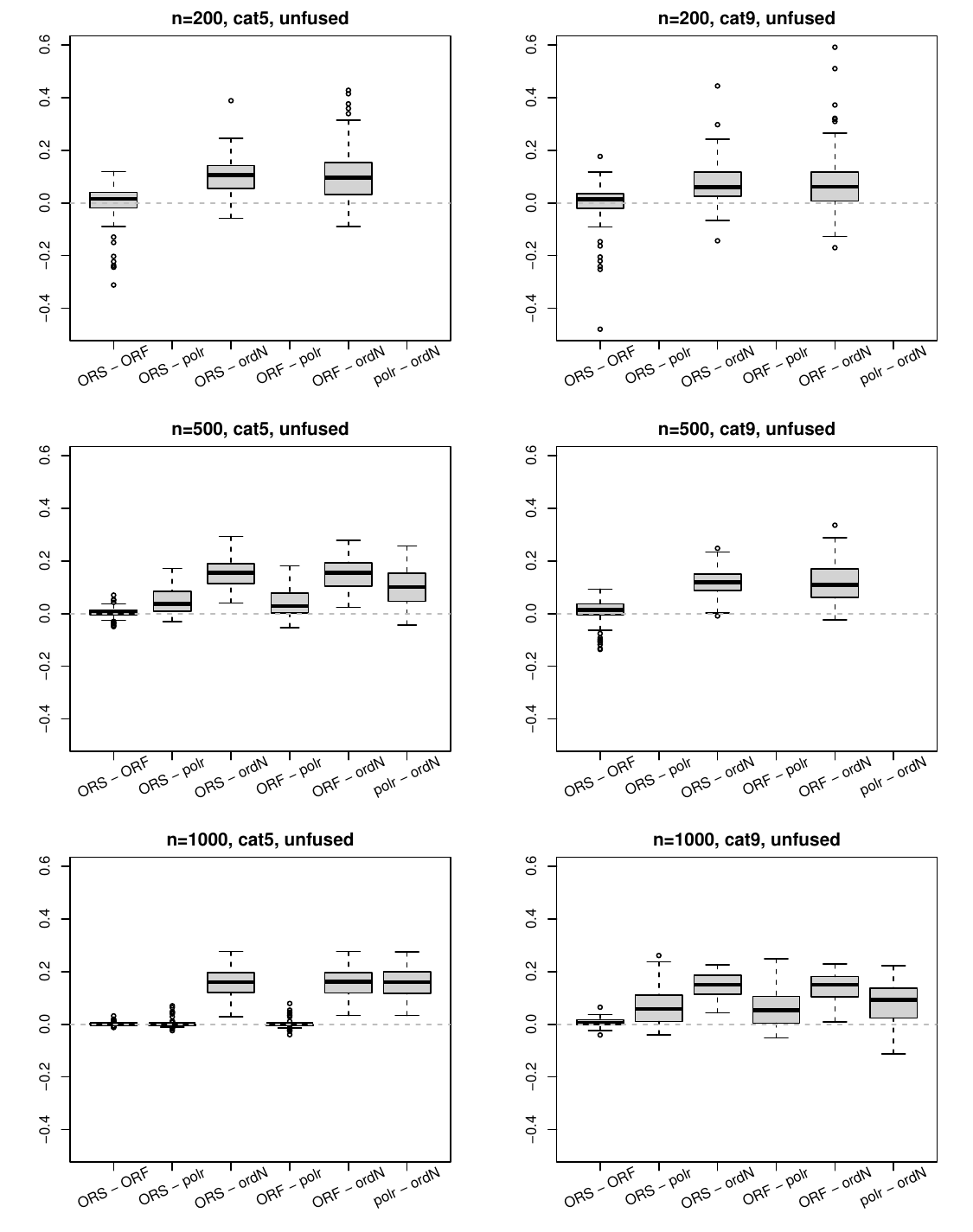} 
 \captionof{figure}{\label{hoshiyar:fig6f} Differences in AUC with true unfused effects. The missing entries correspond to the cases where polr estimation failed. }
 \end{figure}

\begin{figure}[H]\centering
 \includegraphics[width=\linewidth]{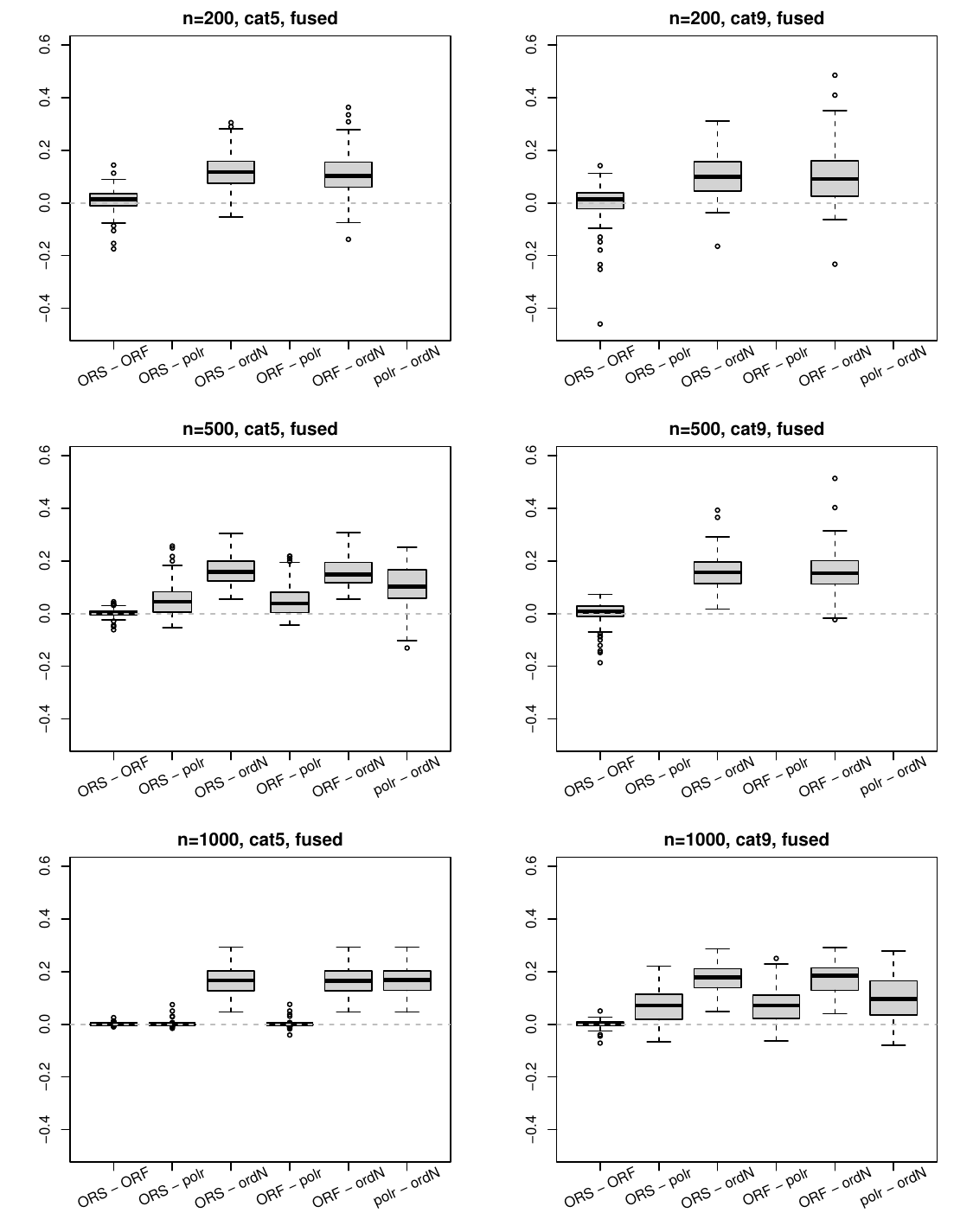} 
 \captionof{figure}{\label{hoshiyar:fig6g} Differences in AUC with true fused effects.  The missing entries correspond to the cases where polr estimation failed. }
 \end{figure}

\begin{figure}[H] 
\centering
\includegraphics[width=\linewidth]{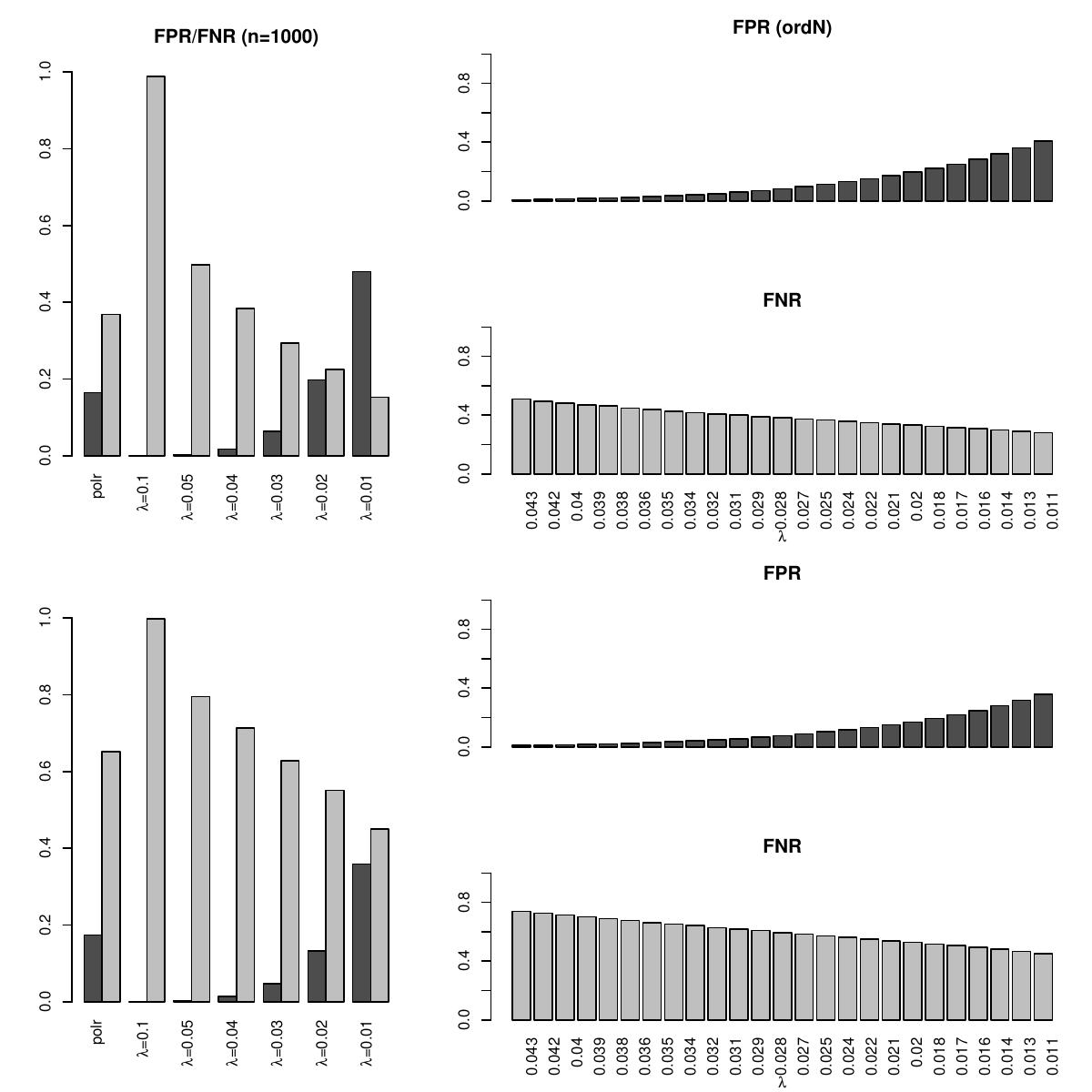}
\caption{\label{Figure_S6} 
Evaluation of polr and ORF: false positive/negative rates (FPR/FNR) concerning identification of relevant differences (i.e., fusion) of dummy coefficients with $n=1000$. Top: $5$ levels with fused effects; bottom: $9$ levels with fused effects.
}
\end{figure}

In this section (Figure~\ref{Figure_S1_1} to \ref{Figure_S6}, and Table~\ref{tab:pvalues}) we show additional simulation results for the setup described in the main paper (Section~$4.1$). 
The ordinal smooth selection and the ordinal fusion penalty are compared to the standard proportional odds model fit by \texttt{polr} and combined with AIC-based forward stepwise selection where the ordinal covariate is treated as nominal, i.e., it is dummy-coded without any penalization, and secondly, to the standard lasso penalty for selection,
treating the predictor as numeric and employing \texttt{ordinalNet}.


\begin{table}[H]
    \centering
        \resizebox{\textwidth}{!}{%
    \begin{tabular}{cccccc ccc}
        \toprule 
        & & & Selection  & Selection  & Selection & Fusion & Fusion & polr \\
  $n$ & no. of categories & effects &  vs. &       vs. &       vs. &        vs. &    vs. &     vs. \\ 
         & & &   Fusion  & polr  & ordinalNet & polr & ordinalNet & ordinalNet \\
        \midrule
        \multirow{6}{*}{200} & \multirow{3}{*}{5} & Fused & $*$ &  & *** & & *** & \\
        & & Unfused & $ 0.5617$ &    &  *** & & *** &\\
        & \multirow{3}{*}{9} & Fused & $0.8083$ &   & ***  & & *** & \\
        & & Unfused & $0.7601$ &   & ***  & & *** & \\
        \midrule
        \multirow{6}{*}{500} & \multirow{3}{*}{5} & Fused & $0.1606$ & *** & ***  & *** & *** & *** \\
        & & Unfused & $ **$ & *** & ***  & *** & *** & *** \\
        & \multirow{3}{*}{9} & Fused & $0.8387$ &   & ***  & & *** & \\
        & & Unfused & $0.1747$ &   & ***   & & *** & \\
        \midrule
        \multirow{6}{*}{1000} & \multirow{3}{*}{5} & Fused & $0.2001$ & $0.3844$ & ***  & 0.7448  & *** & *** \\
        & & Unfused & $**$ & $0.3947$ & ***  & 0.8176 & *** & *** \\
        & \multirow{3}{*}{9} & Fused & $0.6236$ & *** & ***  & *** & *** & *** \\
        & & Unfused & ***  & *** & ***  & *** & *** & *** \\
        \bottomrule 
    \end{tabular}
    }
    \caption{P-values (not adjusted for multiple comparisons) of paired t-tests based on AUC values, with the usual coding * $< 0.05$, ** $< 0.01$, *** $< 0.001$. The missing entries correspond to the cases where \texttt{polr} estimation failed. }
    \label{tab:pvalues}
\end{table}

 
\section{Luxury Food Data: Supplementary Tables and Figures}

\begin{figure}[H] 
\centering
\includegraphics[width=0.65\linewidth]{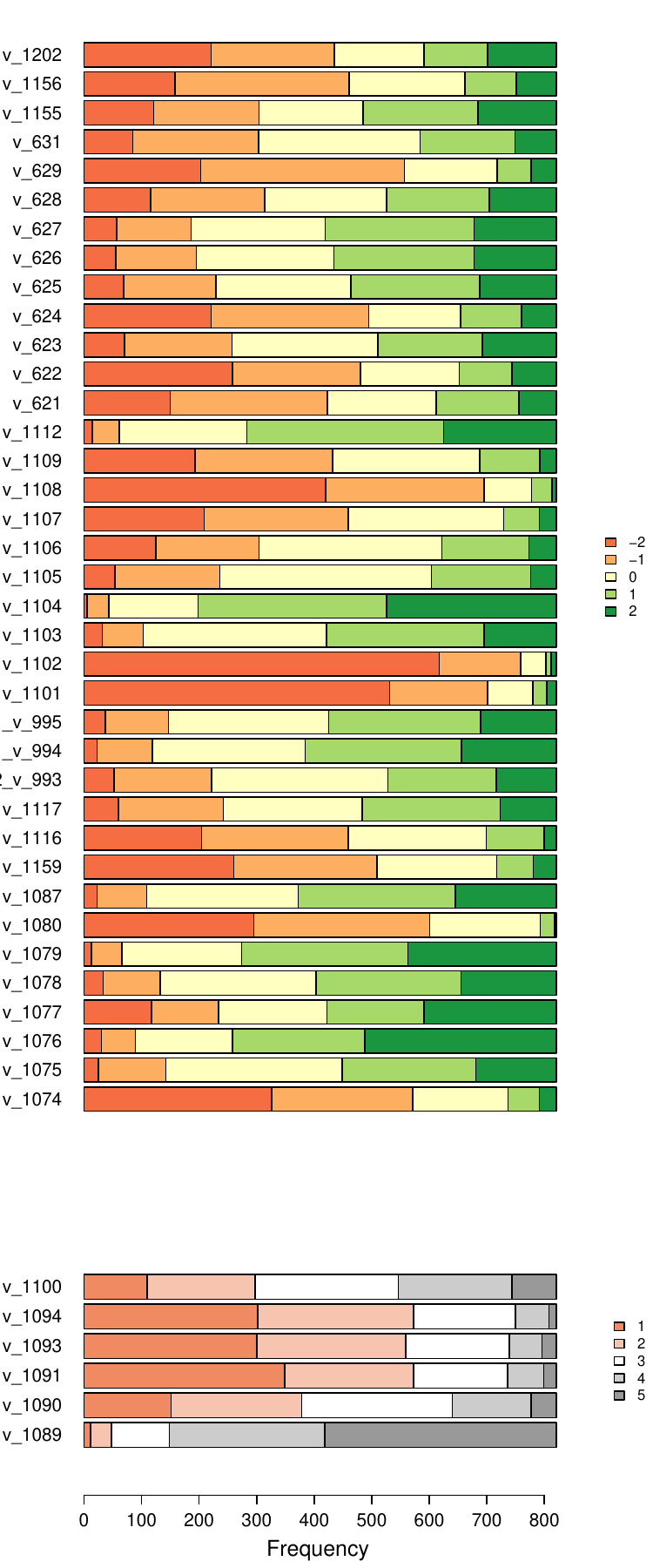}
\caption{\label{Figure_S8_a} 
Summary for items of the luxury food dataset on individual level. Coding scheme for items: $-2$ `strong rejection' to 2 `strong agreement' (top) or 1 `never' to 5 `very often' (bottom).
}
\end{figure}

\begin{figure}[H] 
\centering
\includegraphics[width=\linewidth]{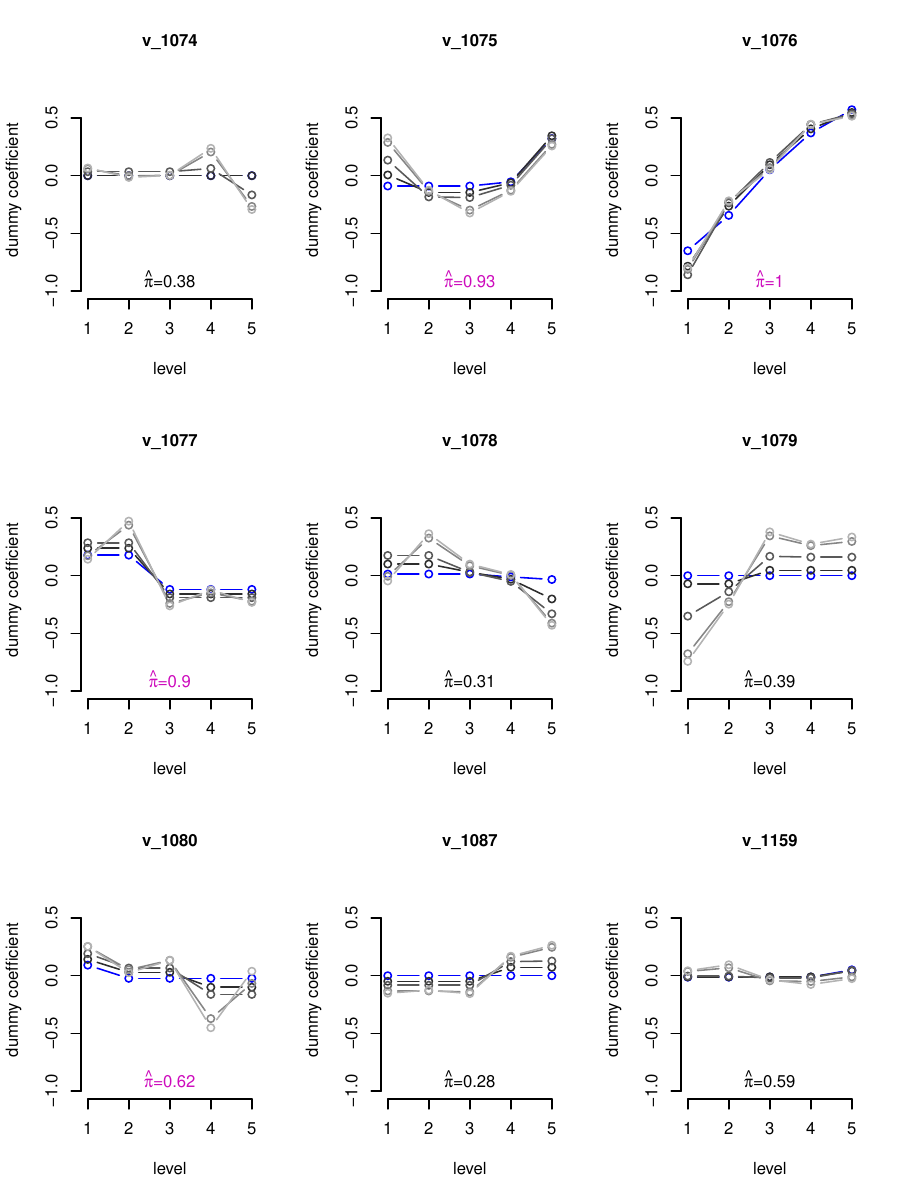}
\caption{\label{Figure_S8_1} 
 Cumulative fused lasso estimates of regression coefficients as functions of class labels ($\lambda \in \{18.5,10,5,1,0.25\}/n$) along with selection frequencies $\hat{\pi}$ according to stability selection. Selected variables according to thresholds $\pi_\text{thr}=0.6$ are highlighted in magenta (where $\lambda$ was set to $15/n$ to achieve comparable results with group lasso). Blue lines correspond to cross-validated/optimal $\lambda=18.5/n$. 
}
\end{figure}

\begin{figure}[H] 
\centering
\includegraphics[width=\linewidth]{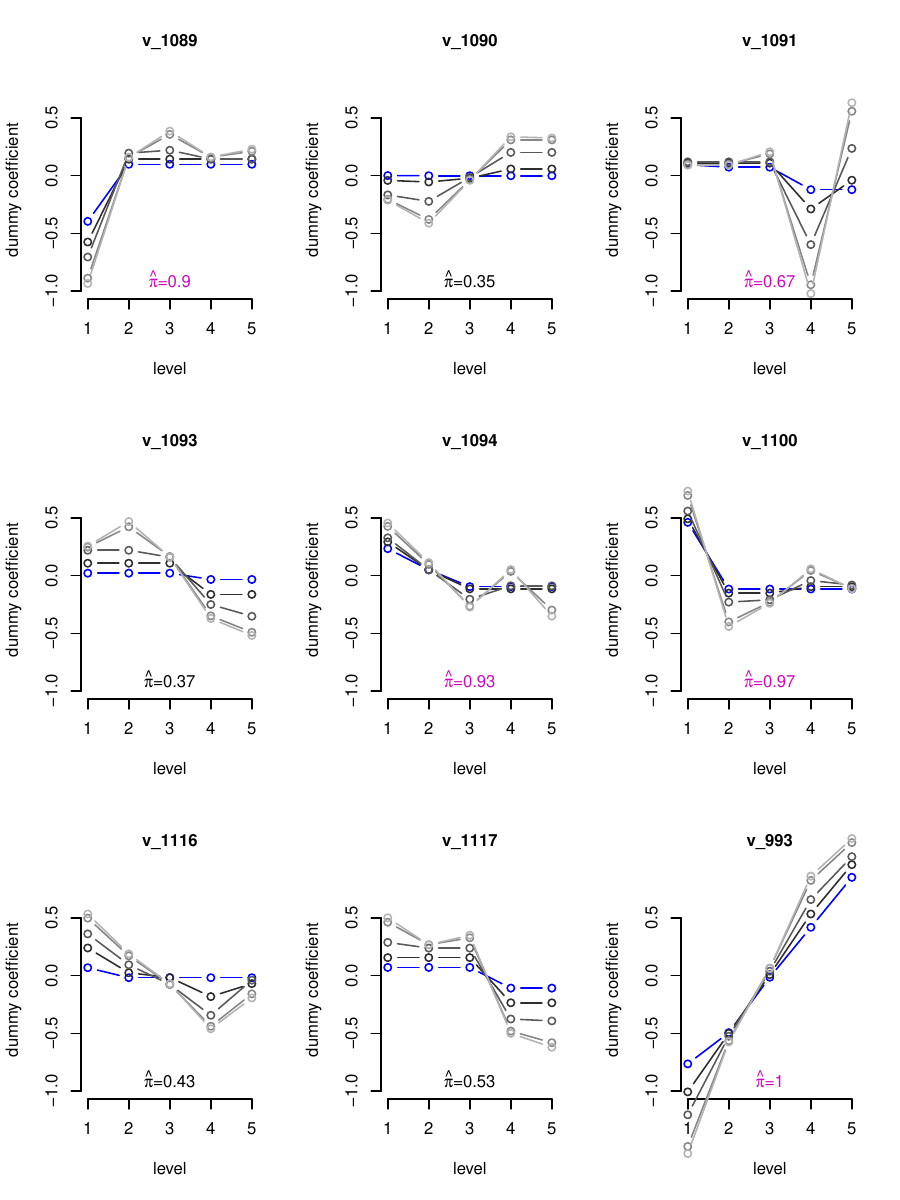}
\caption{\label{Figure_S8_2} 
 Cumulative fused lasso estimates of regression coefficients as functions of class labels ($\lambda \in \{18.5,10,5,1,0.25\}/n$) along with selection frequencies $\hat{\pi}$ according to stability selection. Selected variables according to thresholds $\pi_\text{thr}=0.6$ are highlighted in magenta (where $\lambda$ was set to $15/n$ to achieve comparable results with group lasso). Blue lines correspond to cross-validated/optimal $\lambda=18.5/n$. 
}
\end{figure}

\begin{figure}[H] 
\centering
\includegraphics[width=\linewidth]{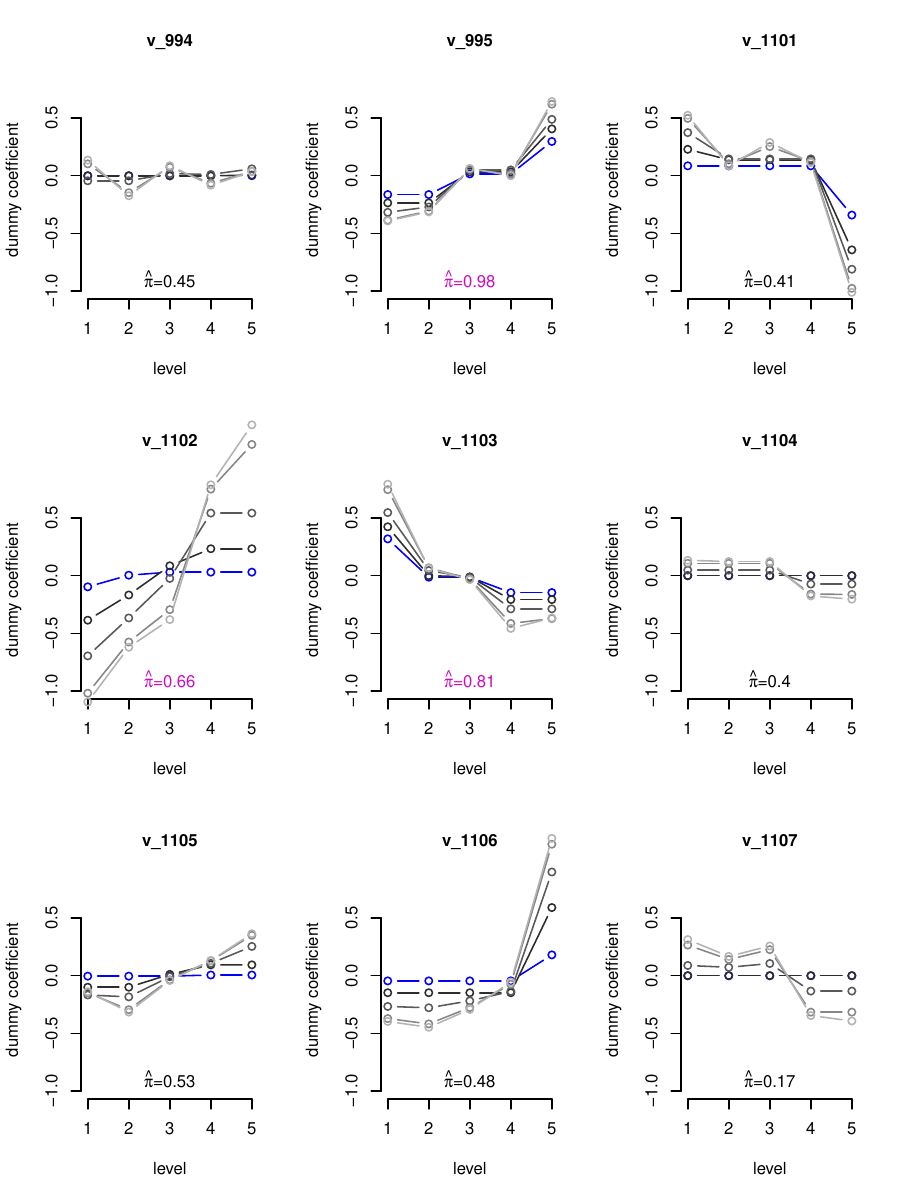}
\caption{\label{Figure_S8_3} 
 Cumulative fused lasso estimates of regression coefficients as functions of class labels ($\lambda \in \{18.5,10,5,1,0.25\}/n$) along with selection frequencies $\hat{\pi}$ according to stability selection. Selected variables according to thresholds $\pi_\text{thr}=0.6$ are highlighted in magenta (where $\lambda$ was set to $15/n$ to achieve comparable results with group lasso). Blue lines correspond to cross-validated/optimal $\lambda=18.5/n$. 
}
\end{figure}

\begin{figure}[H] 
\centering
\includegraphics[width=\linewidth]{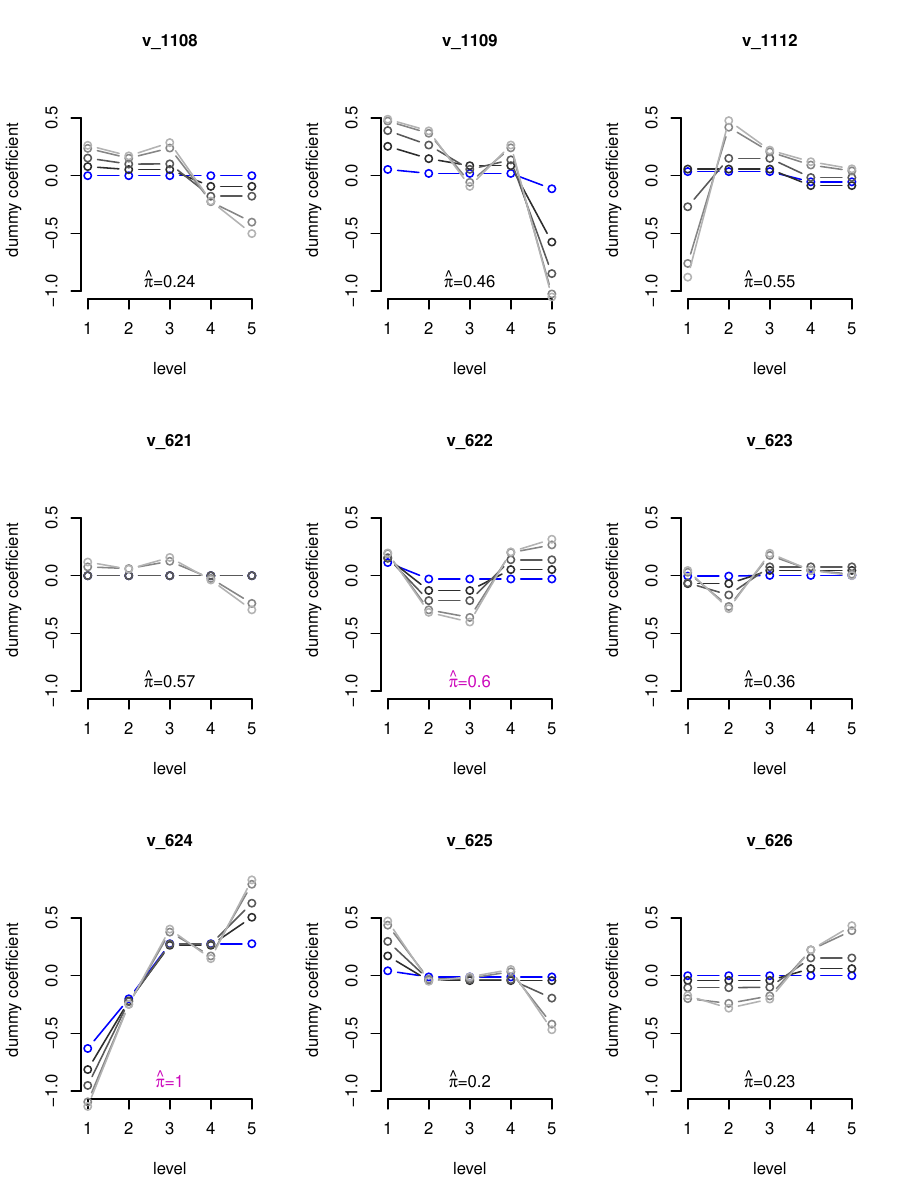}
\caption{\label{Figure_S8_4} 
 Cumulative fused lasso estimates of regression coefficients as functions of class labels ($\lambda \in \{18.5,10,5,1,0.25\}/n$) along with selection frequencies $\hat{\pi}$ according to stability selection. Selected variables according to thresholds $\pi_\text{thr}=0.6$ are highlighted in magenta (where $\lambda$ was set to $15/n$ to achieve comparable results with group lasso). Blue lines correspond to cross-validated/optimal $\lambda=18.5/n$. 
}
\end{figure}

\begin{figure}[H] 
\centering
\includegraphics[width=\linewidth]{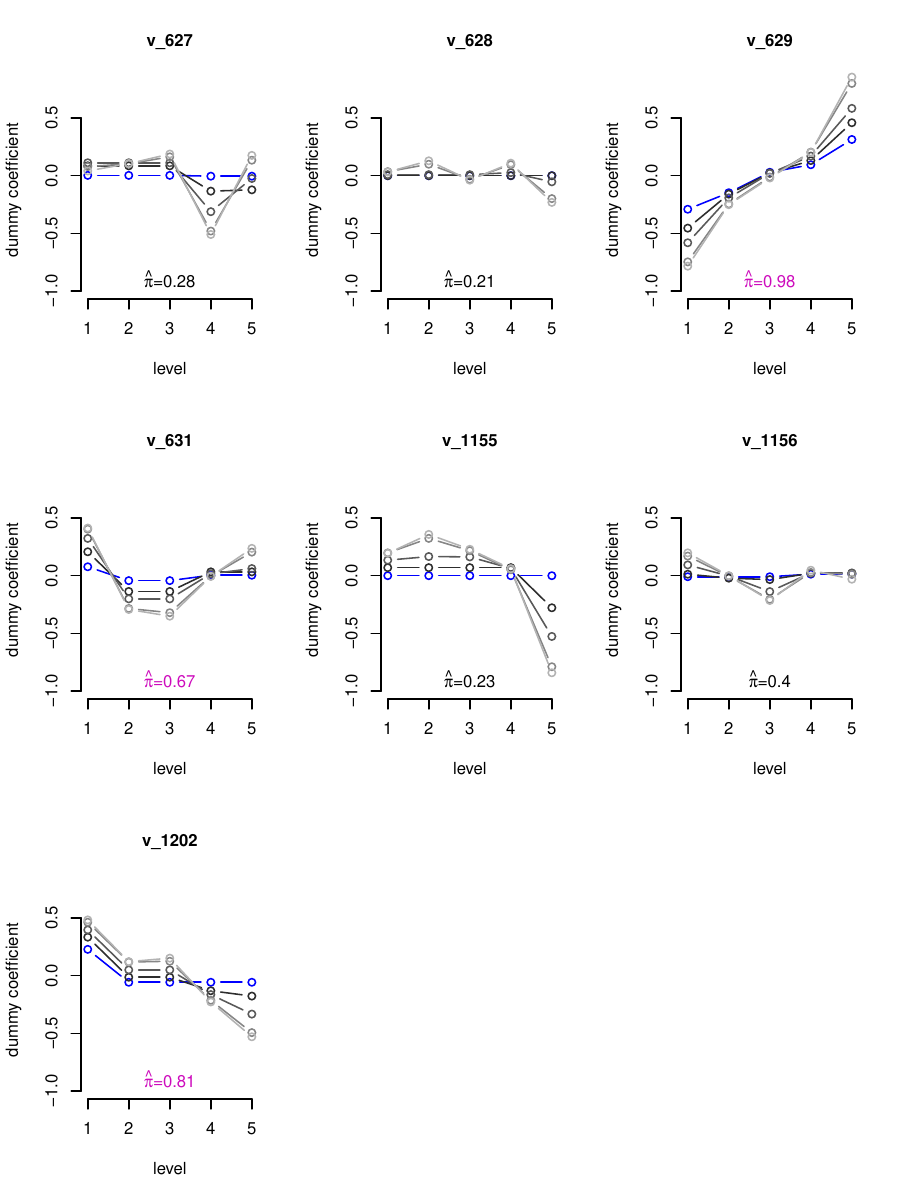}
\caption{\label{Figure_S8_5} 
 Cumulative fused lasso estimates of regression coefficients as functions of class labels ($\lambda \in \{18.5,10,5,1,0.25\}/n$) along with selection frequencies $\hat{\pi}$ according to stability selection. Selected variables according to thresholds $\pi_\text{thr}=0.6$ are highlighted in magenta (where $\lambda$ was set to $15/n$ to achieve comparable results with group lasso). Blue lines correspond to cross-validated/optimal $\lambda=18.5/n$. 
}
\end{figure}

\begin{figure}[H] 
 \centering
 \includegraphics[width=\linewidth]{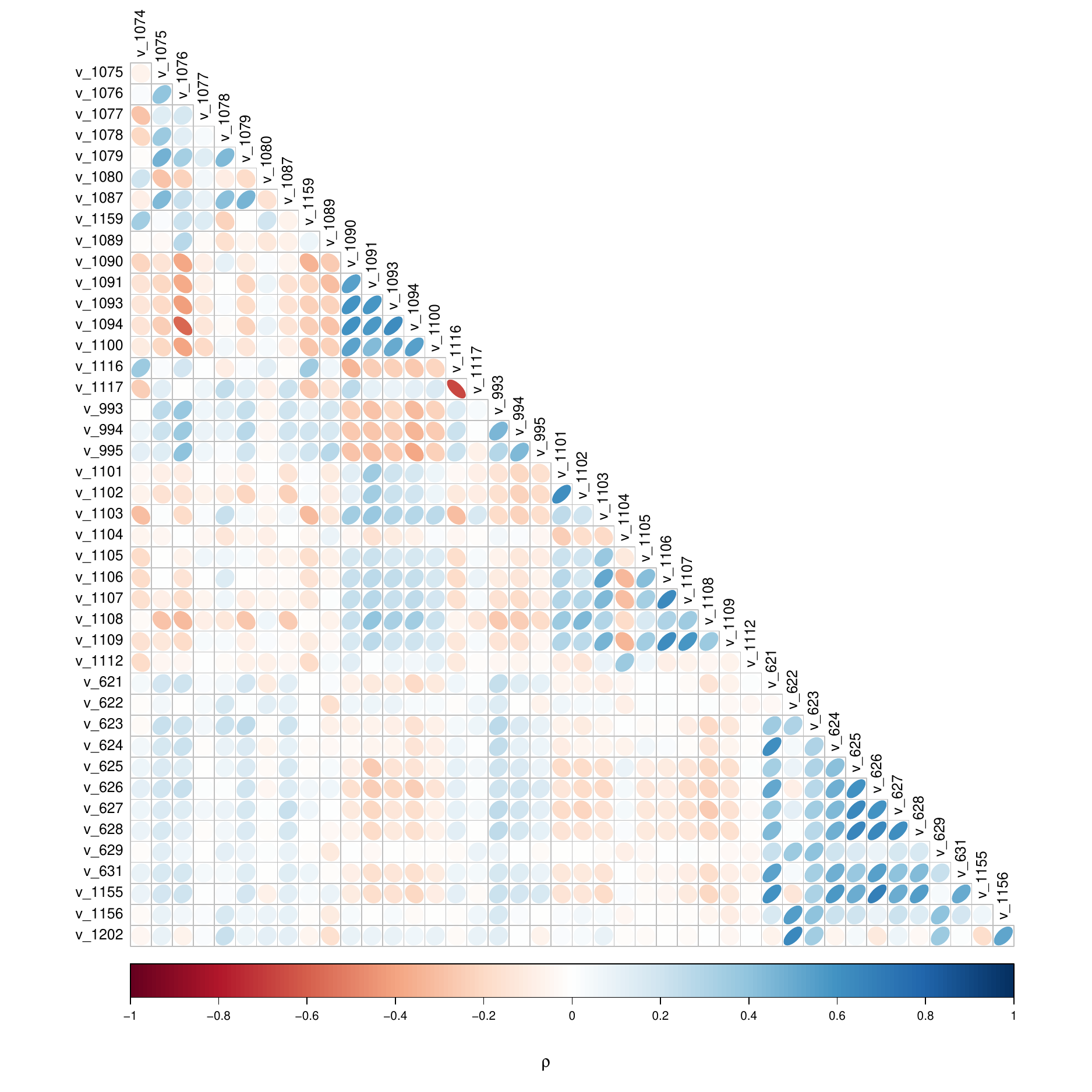}
 \caption{\label{Figure_S8_7} Correlation plot of the luxury food items from Table~\ref{Table_S1}. 
 }
\end{figure}

In this section additional material regarding the luxury food case study is provided. Table~\ref{Table_S1} and Figure~\ref{Figure_S8_a} give a description of the
items along with observed frequencies. Figures~\ref{Figure_S8_1}--\ref{Figure_S8_5} present the cumulative fused lasso estimates of the regression coefficients along with selection frequencies according to stability selection. Figure~\ref{Figure_S8_7} shows the correlation plot of the considered items. 
 
\begin{table}[H]
 \resizebox{0.9\linewidth}{!}{
\centering
\begin{tabular}{llll}
& & & \\ 
  \hline
  & \textbf{Eating habits} &  &   \\ 
  \hline
       &  &      &                 \\
\emph{v\_1074} & I like to cook myself. & &  \\ 
 \emph{v\_1075} & I often go out to eat at inexpensive restaurants, snack bars or cafes. &  & \\ 
 \emph{v\_1076} & I often go out to eat at more expensive restaurants. & & \\ 
 \emph{v\_1077} & Most of the time another family member cooks for me.  &    &   \\ 
 \emph{v\_1078} & I often buy ready-made products.  &   &   \\ 
 \emph{v\_1079} & I often order from a delivery service (e.g. pizza service).   &  &   \\ 
 \emph{v\_1080} & I prefer to eat at home.   &    &   \\ 
 \emph{v\_1087} & I often just take something from the bakery (or similar) instead of big meals.  &   &   \\ 
 \emph{v\_1159} & On weekends, we (my family and I/my friends and I) take a lot of time to eat together. &    &    \vspace{0.25cm} \\   
\vspace{0.25cm} 
Coding: &  $-2=$ `not true at all', $\ldots \,$, $2 = $ `absolutely true' & &   \\
     \hline
       & \textbf{Shopping places: Where do you buy most of your food?}  & &    \\ 
     \hline
       &  &      &                 \\
 \emph{v\_1089} & In discount stores (e.g. Aldi, Lidl, Netto, Penny).   &   &   \\ 
 \emph{v\_1090} & At the weekly market.  &    &   \\ 
 \emph{v\_1091} & In the organice food store (e.g. Alnatura).  &   &   \\ 
 \emph{v\_1093} & In the farm store.  &    &   \\  
 \emph{v\_1094} & In the delicatessen store.  &    &   \\ 
 \emph{v\_1100} & In the specialized trade (e.g. meat, cheese, fruit/vegetable, wine store).  &    &   \vspace{0.25cm} \\   
\vspace{0.25cm} 
Coding: &   $1=$ `never', $\ldots \,$, $5 = $ `very often' & &   \\
     \hline
       & \textbf{Shopping habits (purchasing involvement)}  & &    \\ 
     \hline
       &  &      &                 \\
 \emph{v\_1116} & I like to take my time shopping for groceries.   &    &   \\ 
 \emph{v\_1117} & Food shopping has to be fast for me.   &    &   \vspace{0.25cm} \\   
\vspace{0.25cm} 
Coding: &  $-2=$ `strongly disagree', $\ldots \,$, $2 = $ `fully agree'  & &  \\
     \hline
       & \textbf{What importance does the price of food have for you? (price-value)}  & &    \\ 
     \hline
       &  &      &                 \\ 
 \emph{dupl2\_v\_993} & I associate a high price in food with particularly good quality.   &    &   \\ 
 \emph{dupl1\_v\_994} & I am more likely to buy a certain food item if the price is
comparatively high.  &    &   \\ 
 \emph{dupl1\_v\_995} & When buying food, the price for me is completely undecisive.  &    &    \vspace{0.25cm} \\   
\vspace{0.25cm} 
Coding: &   $-2=$ `strongly disagree', $\ldots \,$, $2 = $ `fully agree'  & &  \\
     \hline
       & \textbf{Nutrition style} & &    \\ 
     \hline
       &  &      &                 \\ 
 \emph{v\_1101} & Vegetarian.  &    &   \\ 
 \emph{v\_1102} & Vegan.  &    &   \\ 
 \emph{v\_1103} & I pay attention to a healthy diet.   &    &    \\  
 \emph{v\_1104} & I eat everything I like. &   &    \\ 
\emph{v\_1105} & I eat in small quantities.  &  &  \\  
\emph{v\_1106} & I eat a low-fat diet. &  &  \\ 
\emph{v\_1107} & I eat a low-carbohydrate diet.  &  &  \\  
\emph{v\_1108} & I eat a functional diet (e.g. sports nutrition). &  &  \\ 
\emph{v\_1109} & I eat slimming.  &  &  \\  
\emph{v\_1112} & I eat a little bit of everything.  &  &   \vspace{0.25cm} \\   
\vspace{0.25cm} 
Coding: & $-2=$ `not true at all', $\ldots \,$, $2 = $ `absolutely true' & &   \\
     \hline
       & \textbf{Luxury food} & &    \\ 
     \hline
       &  &      &                 \\ 
\emph{v\_621} & ...has a particularly fine taste.  &  &  \\  
\emph{v\_622} & ...is particularly expensive.  &  &  \\  
\emph{v\_623} & ...can impress my guests.  &  &  \\  
\emph{v\_624} & ...has a particularly high quality.  &  &  \\  
\emph{v\_625} & ...comes from organic farming.  &  &  \\  
\emph{v\_626} & ...comes from regional cultivation.  &  &  \\  
\emph{v\_627} & ...bears a fair trade seal.  &  &  \\  
\emph{v\_628} & ...it comes from particularly species-appropriate animal husbandry.  &  &  \\  
\emph{v\_629} & ...has a certain rarity value.  &  &  \\  
\emph{v\_631} & ...is a specialty of a country/region.  &  &  \\  
\emph{v\_1155} & ...is particularly fresh.  &  &  \\  
\emph{v\_1156} & ...has an exclusive brand.  &  &  \\  
\emph{v\_1202} & ...tends to be consumed by the rich/better off.  &  &  \vspace{0.25cm} \\   
\vspace{0.25cm} 
Coding: &  $-2=$ `I do not associate with luxury at all', $\ldots \,$, $2 = $ `I strongly associate with luxury' & &   \\
\hline
\end{tabular}
}
\caption{\label{Table_S1} Items (covariates) in the luxury food dataset. 
}
\end{table}


\newgeometry{left=20mm, right=20mm, top=20mm, bottom=20mm}

\newgeometry{left=35mm, right=35mm, top=20mm, bottom=20mm}

\bibliography{sm_references}

\end{document}